\documentclass[fleqn,usenatbib]{mnras}

\usepackage[T1]{fontenc}

\DeclareRobustCommand{\VAN}[3]{#2}
\let\VANthebibliography\thebibliography
\def\thebibliography{\DeclareRobustCommand{\VAN}[3]{##3}\VANthebibliography}

\usepackage{graphicx}	
\usepackage{amsmath}	
\usepackage{amssymb}	
\usepackage{dblfloatfix}
\usepackage{caption}
\usepackage{nccmath}
\usepackage{comment}
\usepackage{multirow}

\usepackage[justification=centering]{caption}

\usepackage{newtxtext,newtxmath}

\defcitealias{afruni19}{AFP19}
\defcitealias{afruni20}{AFP21}

\title[]{Inflow of low-metallicity cool gas in the halo of the Andromeda galaxy}

\author[A. Afruni et al.]{
Andrea Afruni,$^{1}$\thanks{E-mail: afruni@astro.rug.nl}
Gabriele Pezzulli$^{1}$\ \&
Filippo Fraternali$^{1}$
\\
$^{1}$Kapteyn Astronomical Institute, University of Groningen,Landleven 12, 9747 AD Groningen, The Netherlands\\
}

\date{Accepted XXX. Received YYY; in original form ZZZ}

\pubyear{2021}

\begin{document}
\label{firstpage}
\pagerange{\pageref{firstpage}--\pageref{lastpage}}
\maketitle

\begin{abstract}
As the closest $L^{\ast}$ galaxy to our own Milky Way, the Andromeda galaxy (M31) is an ideal laboratory for studies of galaxy evolution. The AMIGA project has recently provided observations of the cool ($T\sim10^4$ K) phase of the circumgalactic medium (CGM) of M31, using HST/COS absorption spectra along $\sim40$ background QSO sightlines, located up to and beyond the galaxy virial radius. Based on these data, and by the means of semi-analytic models and Bayesian inference, we provide here a physical description of the origin and dynamics of the cool CGM of M31. We investigate two competing scenarios, in which (i) the cool gas is mostly produced by supernova(SN)-driven galactic outflows or (ii) it mostly originates from infall of gas from the intergalactic medium. In both cases, we take into account the effect of gravity and hydrodynamical interactions with a hot corona, which has a cosmologically motivated angular momentum. We compare the outputs of our models to the observed covering factor, silicon column density and velocity distribution of the AMIGA absorbers. We find that, to explain the observations, the outflow scenario requires an unphysically large (> 100\%) efficiency for SN feedback. Our infall models, on the other hand, can consistently account for the AMIGA observations and the predicted accretion rate, angular momentum and metallicity are consistent with a cosmological infall from the intergalactic medium.

\end{abstract}




\section{Introduction}\label{intro} 
The Andromeda galaxy (M31) is our closest external $L^{\ast}$ galaxy and therefore represents an ideal laboratory for galaxy evolution studies, in many respects even more than our own Milky Way (MW). The privileged view that we have on this star-forming, disc galaxy, has allowed us to extensively observe and study it in detail for decades, unveiling a large number of its characteristics, including the properties of the stellar \citep[e.g.][]{widrow03,walterbos88,courteau11} and dark matter components \citep[e.g.][]{corbelli10,vandermarel12,zhang21mass}, its current and past star-formation rate \citep[e.g.][]{rahmani16,williams17} and the properties of the different components of its interstellar medium \citep[e.g.][]{carignan06,draine14,calduprimo16,kavanagh20}.\\
While the stellar halo of M31 is well known and has been extensively studied in the past \citep[e.g.][]{ibata01,gilbert18,mcconnachie18,escala20}, the current picture of its gaseous halo is however much less complete. The gaseous environment of M31 has been observed in
neutral hydrogen (HI) emission, but only clouds within a few tens of kpc from
the disk have been detected (in analogy to other nearby star-forming galaxies,
e.g.\ \citealt{sancisi08,putman12}), with the exception of a more extended  `HI bridge’ that connects M31 with its companion galaxy M33 \citep[e.g.][]{braun04,thilker04,lockman12,wolfe16}. There is to date only limited and indirect evidence of the presence of a diffuse hot medium extending up to the galaxy virial radius \citep[][]{zhang21mass,putman21}.\\
Studies on the MW and on external nearby galaxies have shown how galactic halos typically contain a large amount of ionized gas, the circumgalactic medium (CGM), which represents the interface between the intergalactic medium (IGM) and the central galaxy. A significant amount of baryons seems to reside in the CGM \citep[e.g.][]{gatto13,werk14} and, therefore, the role of this medium is believed to be crucial for the growth and evolution of galaxies. This gas can be used to study both the accretion of gas towards the galaxy \citep[e.g.][]{keres09} and the effect of feedback on the galactic environment \citep[e.g.][]{nelson19}. Unveiling the properties and dynamics of this medium is therefore key for our understanding of the cycle of baryons in the halos of galaxies \citep[e.g.][]{ford14}.\\
The CGM is composed of distinct phases with very different temperatures and densities, from the cool ($T\sim10^4$ K, see for example \citealt{borthakur15,keeney17,huang21}) to the hot ($T\sim10^{6-7}$ K, see for example \citealt{anderson11,li17}) phase. While the hot medium is primarily observed in X-ray, the cool gas is generally observed in the UV band, using various absorption lines in the spectra of background quasi-stellar objects (QSOs), typically with only one line-of-sight per galaxy \citep[see][and references therein]{tumlinson17}.
To date, the cool phase of the CGM has often been associated both with inflow of external gas \citep[e.g.][]{bouche13,chen19} and outflows from the central galaxy \citep[e.g.][]{rubin14,schroetter19}, but what is the relative importance of these two processes and which one is dominant in determining the properties of this gas is still unclear and debated.\\
While our position inside the galactic
disc limits the study of the extended circumgalactic gas in the MW \citep[see][]{zheng15}, M31 represents an ideal candidate for the study of this medium.
Recently, a new set of observations from the AMIGA project \citep{lehner15,lehner20p} has shed light on the cool ($T\sim10^4$ K) and warm ($T\gtrsim10^5$ K) phases of the CGM of M31. In particular, they analysed the absorption lines of a number of metal ions, including three different ionization states of Silicon (Si$\,$II, Si$\,$III and Si$\,$VI), in the spectra of more than 40 background QSO sightlines, spanning up to and beyond the galaxy virial radius. These data represent a unique and unprecedented view on the CGM of a single galaxy, showing the distribution of this cool gas across the entire galaxy halo, in contrast with typically one single line of sight available through the halos of more distant galaxies \citep[but see][]{lopez18}. However, a clear physical picture able to interpret these observations is still missing. The aim of this paper is to fill this gap and, building on this unique dataset, to provide a physical model for the origin and dynamics of the CGM of M31.\\
To describe the dynamics and properties of the CGM of M31, we adopted a similar approach as in our previous works (\citealt{afruni19,afruni20}, hereafter \citetalias{afruni19,afruni20}), where we made use of semi-analytical parametric models to interpret observations of the CGM of low-redshift galaxies. In particular, we describe the CGM as formed by a population of cool clouds flowing through the galactic halo and embedded in a hot coronal gas. We then compare the predictions of these models with real data through a Bayesian analysis in order to find the best physical parameters that describe the observations. In \citetalias{afruni19} and \citetalias{afruni20} we have coupled our models with observational data from two different samples of respectively early-type \citep[ETGs][]{chen18,zahedy19} and star-forming \citep[see][]{werk13,borthakur15} galaxies. We have found that in the ETGs the cool CGM is well reproduced by an inflow of clouds falling at a rate consistent with the accretion predicted by cosmological models; for star-forming galaxies we investigated instead the impact of galactic feedback on the surrounding CGM, finding that supernova driven galactic winds cannot reproduce the cool absorbers observed at distances of hundreds of kpc from the central galaxy, given the unphysical energy requirements that this scenario would imply.\\
Our semi-analytic approach allows therefore a physically motivated description of the dynamics of the CGM clouds throughout the galaxy halo, taking into account the effects of gravity, pressure and the drag force acted by the hot medium. These studies are then complementary to cosmological hydrodynamical simulations, which are still limited by insufficient resolution in the circumgalactic medium \citep[e.g.][]{vandevoort19,peeples19}.\\
Having only one line-of-sight per galaxy, in our previous studies we have assumed, in both samples of star-forming and early-type galaxies, that the properties of the CGM would be statistically similar for all the galaxies of the same type.
The AMIGA data allow us instead to go beyond this statistical approach and to study in detail the CGM of a single galaxy. Moreover, the measures of the total silicon column densities, available from this survey for each line of sight, will allow us to directly estimate the metallicity of the cool CGM, a property that is still very uncertain and debated in the literature \citep[see][]{prochaska17,wotta19,pointon19}. This is fundamental to constrain the origin of this medium, given that finding low-metallicity gas would point at IGM accretion, while metallicities close to the solar value would imply gas ejected from the interstellar medium.\\
Both our previous works hint towards a picture where the accretion of external gas is likely the origin of the majority of the cool CGM of galaxies in the local Universe. 
Here, we explore both scenarios of inflow from the intergalactic medium (IGM) and outflow from a galactic wind as a possible description of the cool CGM of M31, taking into account also the rotation of both the hot and the cool phase.\\
In Section~\ref{Observations} we briefly report the observations of the AMIGA project and the data selected for our work; in Section~\ref{model} we describe the semi-analytical parametric models and the way we compare our synthetic observations wit the AMIGA data; in Section~\ref{results} we show the results of our analysis and finally in Sections~\ref{discussion} and \ref{conclusions} we respectively discuss our findings and we report our conclusions. 
\section{Data: the AMIGA Project}\label{Observations}
\begin{figure}
   \includegraphics[clip, trim={0.5cm 0.1cm 1.6cm 0.5cm}, width=\linewidth]{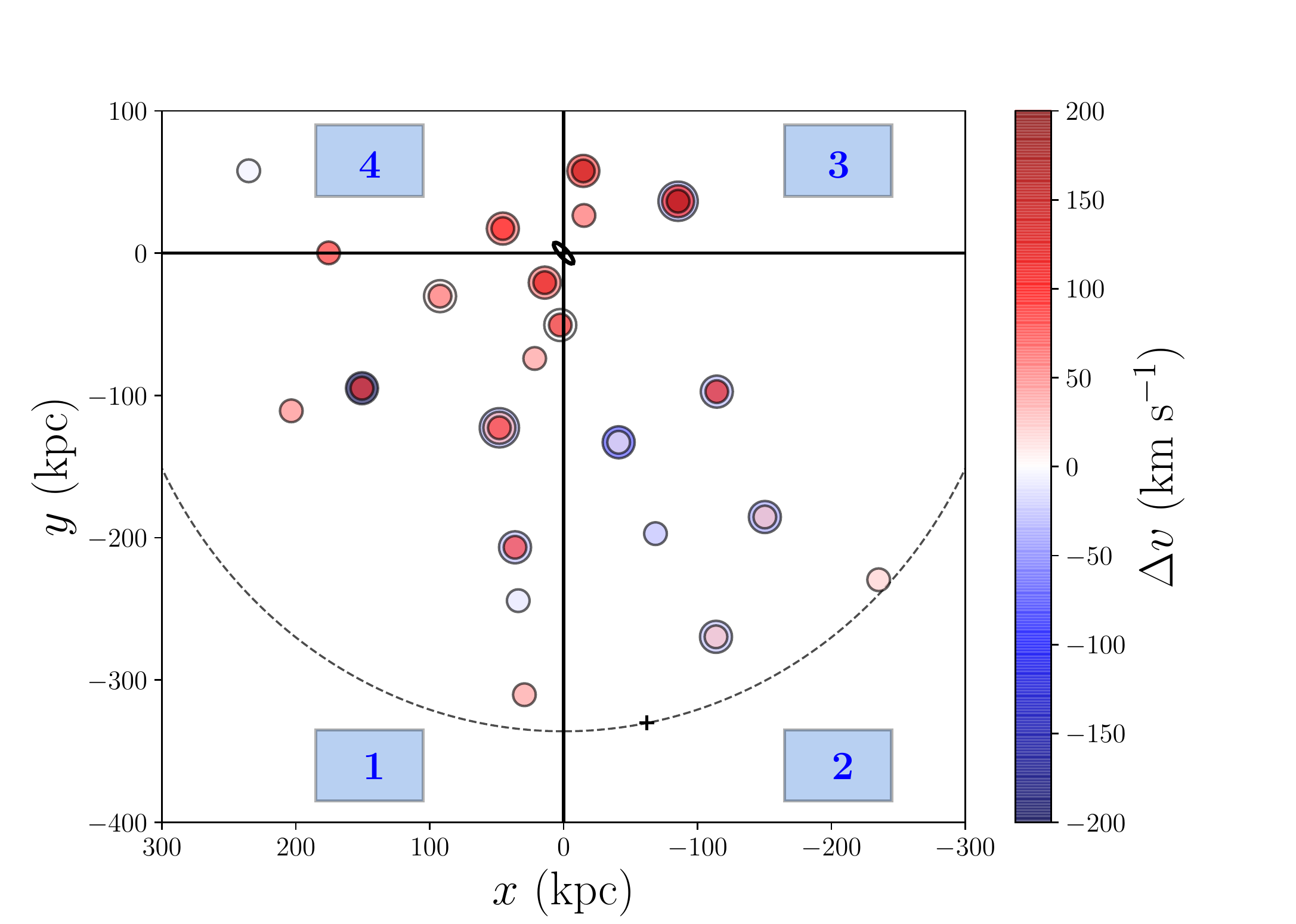}
   \includegraphics[clip, trim={0.5cm 0.1cm 1.6cm 0.5cm}, width=\linewidth]{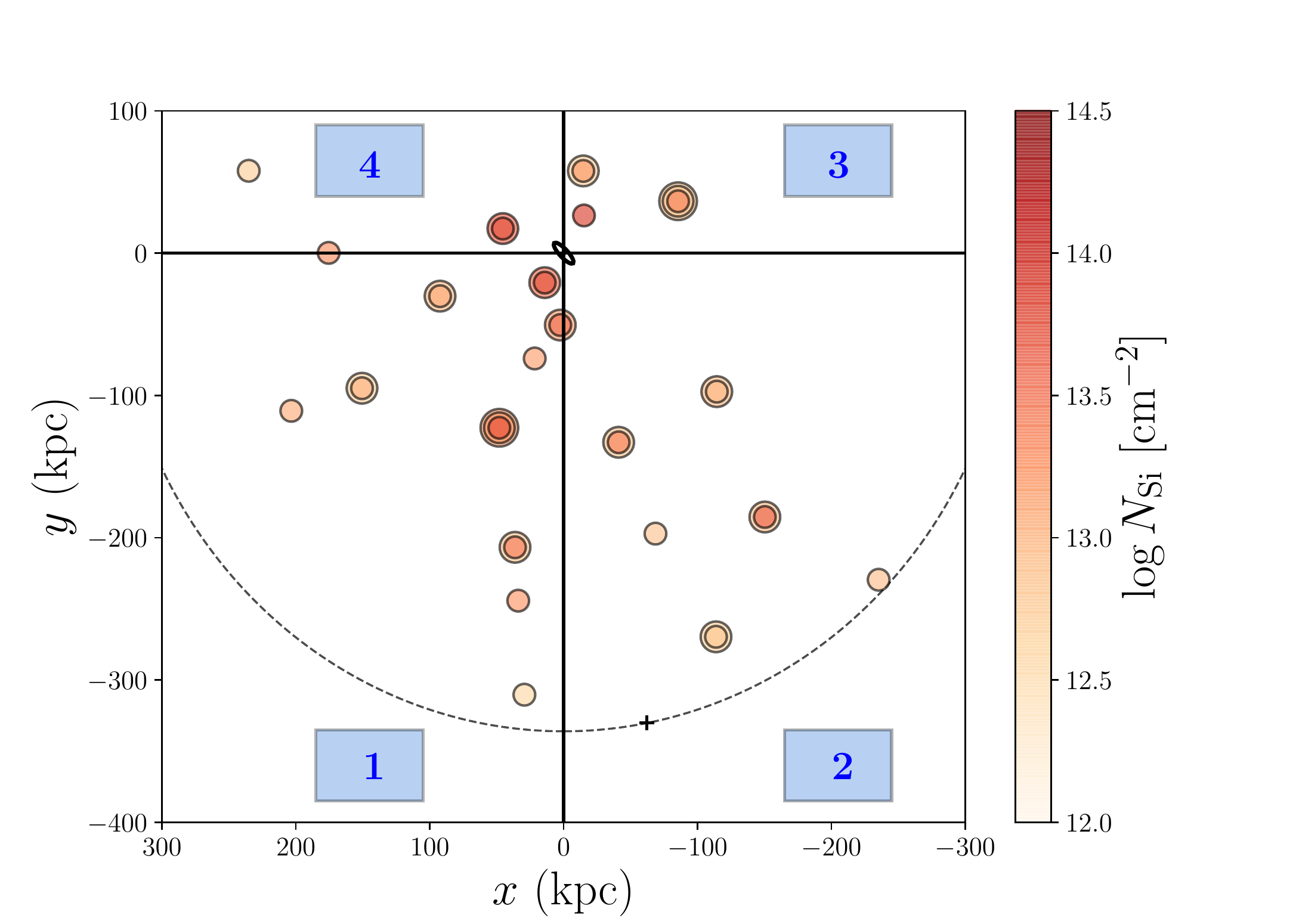}
   \caption{Observational data from the AMIGA project \citep{lehner20p} used in this work. Top: SiIII velocities with respect to the systemic velocity of M31; bottom: total silicon column densities. In both panels the central ellipse represents the disc of M31, with a position angle in this reference frame of $45^{\circ}$, while the circles represent the 23 QSO sightlines at their projected position. Multiple velocity components are shown by overlapping circles with different colors. The cross depicts the only non-detection of SiIII present in our sample. The dashed circle represents the virial radius of M31, while the vertical and horizontal lines are used to divide the plane in 4 quadrants.}
              \label{fig:Obs}%
    \end{figure}

The AMIGA project \citep{lehner15,lehner20p} is an HST large program aiming to characterize the properties of the cool/warm phase of the CGM in the halo of M31, using a sample of 43 QSO sightlines at projected distances from 25 to 569 kpc from the centre of M31. The target QSOs are observed using the COS spectrograph \citep{froning09} with the G130M and/or G160M gratings. 
Here we use in particular the data reported in \cite{lehner20p}, who analysed the cool and warm (temperatures of about $10^{4-5}$ K) CGM of M31 through the study of the absortion lines coming from species such as C$\,$II, Si$\,$II, Si$\,$III, Si$\,$IV and O$\,$VI.
For our analysis, we decided to focus on the absorbers that lie (in projection) within the virial radius of M31 (which we estimate to be 336 kpc, see Section~\ref{model} and Table~\ref{tab:M31prop}). 
Absorbers at distances larger than 350 kpc from the galactic disc have a higher probability of not being directly related to M31 and to possibly be part of a local group medium \citep[see][]{putman21,qu21} \footnote{We have verified that applying our models to the whole data sample of the AMIGA Project has a negligible impact on the findings of this work.}. This selection leaves us with 23 lines of sight (see Figure~\ref{fig:Obs}) out of the 43 of the original AMIGA sample (13 out of the 20 excluded sightlines are non-detections).\\ 
The spectra were inspected in the velocity range $-700 \leq \varv_{\rm{LSR}} \leq -150\ \rm{km}\ \rm{s}^{-1} $. The lower limit corresponds to about 100 km $\rm{s}^{-1}$ less than the most negative velocities associated with the M31 rotation ($\sim-600\ \rm{km}\ \rm{s}^{-1}$ in the LSR, with the systemic velocity of M31 being $-300\ \rm{km}\ \rm{s}^{-1}$, see \citealt{chemin09}), while the upper limit is set to avoid the absorption of the MW, which dominates in the $\varv_{\rm{LSR}}$ ranging from -150 to +50 km $\rm{s}^{-1}$. In addition to the Milky Way, another important source of contamination is the Magellanic Stream (MS), since its line-of-sight velocities can overlap with those expected from the CGM of M31. As discussed in detail in \cite{lehner15,lehner20p}, this contamination was removed from the AMIGA data by excluding, for each line of sight, all the kinematic components with velocities that lie within the MS velocity range, inferred by \cite{Nidever10} using the HI 21 cm emission detected with the Green Bank Telescope. The same cut will be also applied to our models (see Section~\ref{likelihood}).
   \\
Most of the sightlines host multiple kinematic components and the velocities and column densities for each component were then obtained using the apparent optical depth \citep[AOD;][]{savage1991} method. 
For the purpose of this work, we considered the silicon absorption lines as tracers of the cool circumgalactic medium. In particular, following \cite{lehner20p}, we identify the different kinematic components using the Si$\,$III lines, being the Si$\,$III the species with the strongest observable transition. As reported in \cite{lehner20p}, there is a good agreement in the kinematic pattern resulting from low ion absorptions (SiII, CII, SiIII), presumably tracing the same cool gas phase at $T\sim \rm{few}\times 10^{4}$ K. Moreover, detecting the absorption of SiII, Si$\,$III and Si$\,$IV, a direct estimate of the total silicon column density\footnote{In those cases where SiII and/or SiIV were not detected, but SiIII was, we adopted the upper limit for the non-detected lines. This procedure affects the total column densities in about 20$\%$ of the sample.} in the cool
CGM of M31 is possible, largely insensitive to the details of photo-ionization modelling \citep[see][]{lehner20p}. 
The constraint given by the observed total silicon column density will allow us to give an estimate of the gas metallicity directly through the comparison of these densities with the outputs of our models (see Sections \ref{model} and \ref{results}).\\
In the two panels of Figure~\ref{fig:Obs} we show the velocities (top) and the total silicon column densities (bottom) that we will use as constraints for our models. We show in particular the cool gas kinematics with respect to the systemic velocity of M31, obtained from the observed LSR velocities using the transformation reported in \cite{lehner20p}. 
\section{Model}\label{model}
The model used in this work has many similarities with the models presented in \citetalias{afruni19} and \citetalias{afruni20}. As in our previous works, we describe the cool CGM as a population of clouds in pressure equilibrium with the ambient hot medium (corona), flowing towards the galaxy or outflowing from the central disc. The cool CGM is subject to the gravitational force of the galaxy and the DM halo and to the interaction with the corona. To describe the cloud motion we use the python package GALPY \citep{bovy15}, which we modified in order to take into account the drag force acted by the hot coronal gas (see \citealt{fraternali08}, \citetalias{afruni20}). We explore two separate scenarios of outflow and inflow as an origin of the cool clouds, each defined by a different set of 5 free parameters, as we explain in detail in Section~\ref{coolCGMprop}. 
\subsection{Creation of the M31 setup}\label{M31setup}
\begin{figure}
   \includegraphics[clip, trim={0.2cm 0.2cm 0.8cm 0.4cm}, width=\linewidth]{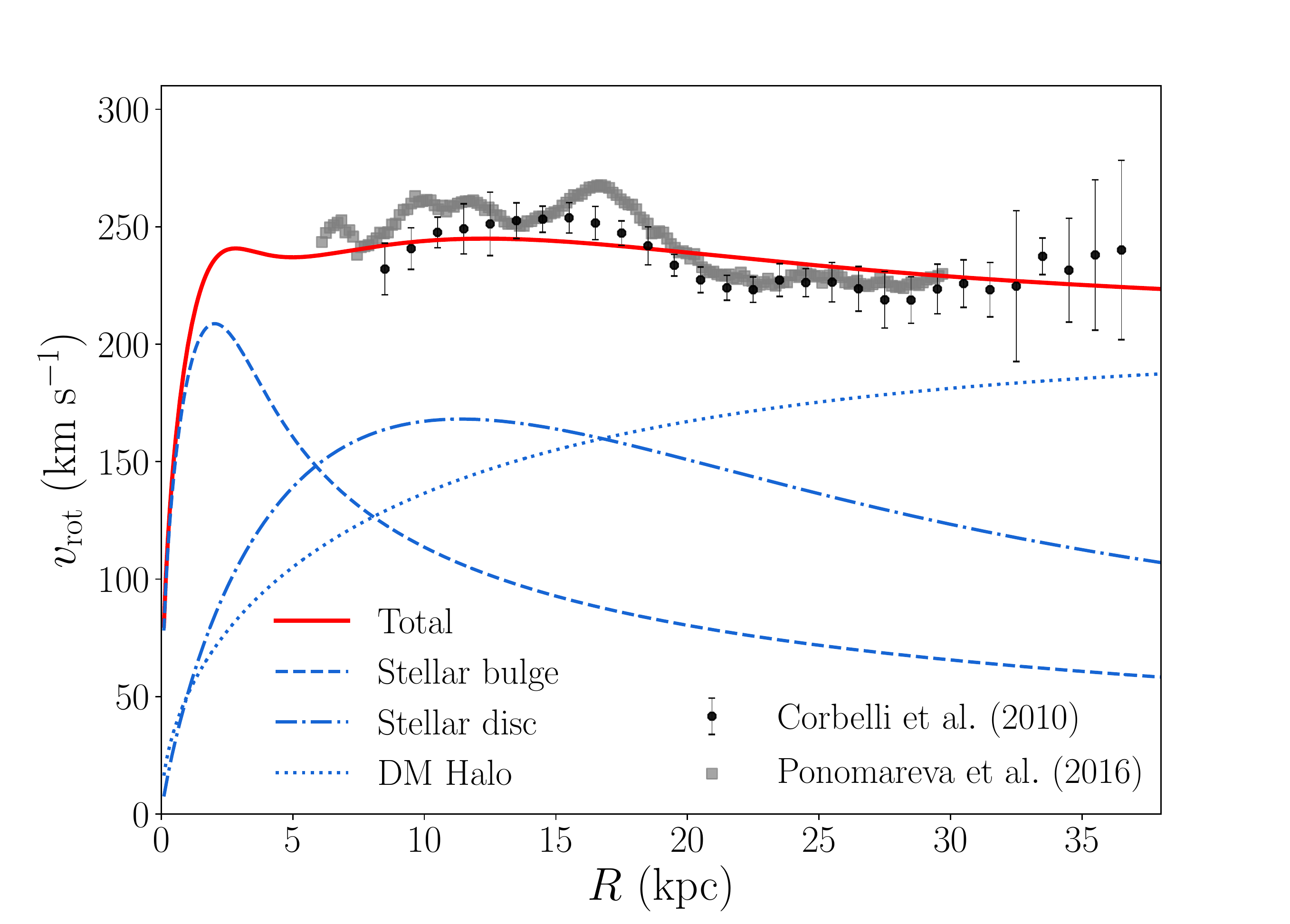}
   \caption{Rotation curve decomposition of M31 adopted in this work. The data points come from HI 21 cm emission and are taken from \protect\cite{corbelli10} and \protect\cite{ponomareva16}. The red curve shows the total circular speed predicted by our model, fairly in agreement with the observations. The total circular speed is given by the sum of three different components: a stellar bulge (dashed curve), a stellar disc (dotted-dashed curve) and a NFW dark matter halo (dotted curve) (see Section~\ref{M31setup}).}
              \label{fig:Rotcurv}%
    \end{figure}

{ 
\renewcommand{\arraystretch}{1.5}
 \begin{table}
\begin{center}
\begin{tabular}{*{3}{c}}
\hline  
\hline
 M31 property & Value & Units\\
\hline 
$M_{\rm{vir}}$ & $2 \times 10^{12}$ & $\rm{M}_{\odot}$\\
$c$ & 10.25& -\\
$r_{\rm{vir}}$ &  336 & kpc \\
$M_{\ast}$& $1.2\times 10^{11}\ $&$M_{\odot}$ \\
$R_{\rm{d}}$& 5.3 & kpc\\
$B/T$&0.25&-\\
$i$& $77$& degrees\\ 
SFR& 0.35 &$M_{\odot}\ \rm{yr}^{-1}$\\
\hline
\end{tabular}
\end{center}
\captionsetup{justification=centering}
\caption[]{Main properties of M31 used in this work.}\label{tab:M31prop}
   \end{table}
   }
To create the setup for our models, we first need to set up the gravitational potential generated by the central galaxy and the dark matter (DM) halo. For the dark matter component we choose a Navarro Frenk White \citep[NFW][]{nfw96} profile, while for the stellar component we adopt a thin exponential disk plus a spherical central bulge with a density distribution equal to
\begin{ceqn}
\begin{equation}\label{eq:bulge}
\rho_{\rm{bulge}}(r)=\rho_{\rm{bulge},0}\left(\frac{r_0}{r}\right)^{\alpha}\exp\left[{-(r/r_{c})^2}\right] ,
\end{equation}
\end{ceqn}
also used by \cite{bovy15} to describe the bulge of the Milky Way and where $\rho_{\rm{bulge},0}$ is the bulge mass density at the reference radius $r_0$.
We adopted a disc scale length $R_{\rm{d}}=5.3$ kpc \citep[see][]{walterbos88,courteau11} and a bulge to total ratio $B/T=0.25$ \citep[][]{widrow03,chemin09}. The mass of both the stellar and the dark matter components are still uncertain, with values in the range $1-1.5\times10^{11}\ M_{\odot}$ for the stellar mass \citep[e.g.][]{corbelli10,sick15,williams17} and $0.6-2.4\times10^{12}\ M_{\odot}$ for the dark matter mass \citep[see][and references therein]{zhang21mass}. In this work, we used intermediate values of $M_{\ast}=1.2\times 10^{11}\ M_{\odot}$ and $M_{\rm{vir}}=2\times 10^{12}\ M_{\odot}$. The choice of these values produces a circular speed that well matches the observed rotation curve of M31 \citep{corbelli10,ponomareva16}, as shown in Figure~\ref{fig:Rotcurv}. We also chose $\alpha=1.2$ and $r_{\rm{c}}=2$ kpc for the bulge profile (see equation \eqref{eq:bulge}) in order to reproduce the observed rotation curve. Note however how the central parts of this curve are not sampled by the observational data, therefore a slightly different choice for $\alpha$ and $r_{\rm{c}}$ is possible. This however would have minimal effects on our final results. 
The adopted virial mass
implies a concentration $c=10.25$ \citep[from][]{dutton14} and a virial radius
\begin{ceqn}
\begin{equation}\label{eq:Rvir}
r_{\rm{vir}}=\left(\frac{2GM_{\rm{vir}}}{\Delta H_0^2}\right)^{1/3}=336\ \rm{kpc}\ ,
\end{equation}
\end{ceqn}
where $H_0=67.4\ \rm{km}\ \rm{s}^{-1}\ \rm{Mpc}^{-1}$ \citep[][]{planck20} and $\Delta=101$ \citep[calculated using the prescription of][]{bryan98}.\\
The galactic disc of M31 has an inclination $i=77^{\circ}$ \citep[see][]{walterbos88,athanassoula06}. The far-side of the galaxy is on the South-East and the near-side on the North-West \citep[see][]{carignan07,chemin09}. We set the intrinsic reference frame of all our models in order to have a positive tangential velocity $\varv_{\rm{t}}$ in the same sense of rotation of M31. All the properties of the Andromeda galaxy used in this work are reported in Table~\ref{tab:M31prop}.
   \begin{figure*}
   \includegraphics[clip, trim={3cm 2.5cm 2cm 4cm}, width=\linewidth]{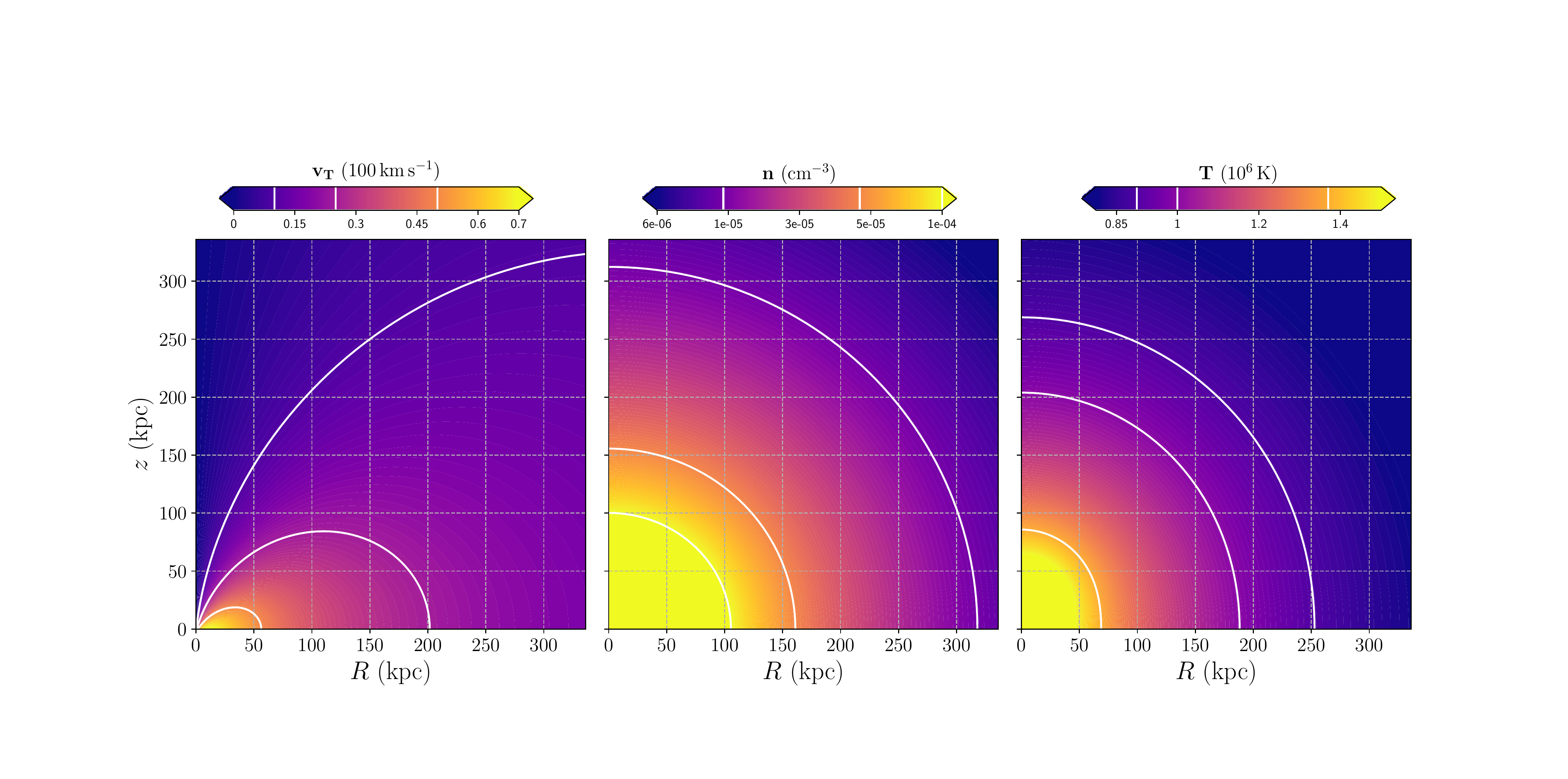}
   \caption{Tangential velocity (left), number density (centre), and temperature (right) of the model of the rotating hot corona described in Section~\ref{model}, as a function of height and cylindrical radius. The model is defined to contain 20\% of the nominal baryons and an angular momentum comparable to that of the DM halo (see details in Appendix~\ref{hotgas}). In each panel, contours are placed at the values indicated on the corresponding colorbar.}
              \label{fig:Corona}%
    \end{figure*}
\subsection{Properties of the hot CGM}\label{hotgasMod}
As in our previous works, we assume that the halo of M31 is filled with a stratified hot corona that extends up to very large distances from the centre. From both observational evidence \citep{hodges16} and theoretical arguments \citep{pezzulli17} we expect the corona of star-forming galaxies to rotate, although less than the local circular speed because of partial pressure support against gravity. For this work we created, using a modified version of the software COROPY from \cite{sormani18}, an axisymmetric model of a corona rotating in the same direction of the galactic disc, in equilibrium with the dark matter halo and with a physically motivated angular momentum. The details of how this model has been constructed can be found in Appendix~\ref{hotgas}.\\
We chose parameters such that the average angular momentum of the hot medium is $l_{\rm{cor}}=3207\  \rm{km}\ \rm{s}^{-1}\ \rm{kpc}$. This is comparable to the specific angular momentum expected for the dark matter halo $l_{\rm{DM}}=2619\  \rm{km}\ \rm{s}^{-1}\ \rm{kpc}$ \citep{cimatti19}. For comparison, the stellar angular momentum\footnote{We neglected the asymmetric drift and we assumed that the bulge is either not rotating ($l_{\rm{M31}}= 1885\  \rm{km}\ \rm{s}^{-1}\ \rm{kpc}$) or rotating at the same rotational velocity of the disc ($l_{\rm{M31}}= 1959\  \rm{km}\ \rm{s}^{-1}\ \rm{kpc}$).} of M31 (estimated following equation 1 in \citealt{pavel21}) is $l_{\rm{M31}}\simeq1900\  \rm{km}\ \rm{s}^{-1}\ \rm{kpc}$ and, as expected, is smaller than (but comparable to) that of the rotating corona \citep[see][and references therein]{pezzulli17}.
We also checked the stability of our model using the Solberg-H{\o}iland criteria \citep{tassoul00,sormani18} and we checked that the cooling time of the hot medium is significantly higher than 2.5 Gyr (the typical timescale of the cool CGM cloud orbits, see Section~\ref{resultsIn}) everywhere in the halo, except for the regions close to the disc plane, which are however not relevant for the findings of this work (see Sections~\ref{resultsIn} and \ref{gasfate}).\\
The mass of the corona in our fiducial model is $\sim6\times10^{10}\ M_{\odot}$ and accounts for $20\%$ of the total baryonic mass associated with the halo of M31. Note that the stellar component of M31 represents almost $40\%$ of the baryonic mass, therefore in our model we have, without considering the cool circumgalactic gas, around $60\%$ of the expected baryons. The rotation velocity, density and temperature of the hot gas are shown in the three panels of Figure~\ref{fig:Corona}.
The corona is rotating significantly in the inner regions, while the rotation velocity declines substantially at large distances from M31. Even though we do not have direct observational estimates of the properties of the corona of M31, our inferred profiles are in broad agreement with current observational estimates of the MW corona \citep[e.g.][]{salem15,hodges16} and are consistent with the ones that we used for other low-redshift star-forming galaxies (\citetalias{afruni20}).\\
Note that in this work we adopt a physically motivated model for the hot CGM, but we have verified that using a simpler, non-rotating model for the corona would not strongly influence our results reported in Section~\ref{results}.
\subsection{Properties of the cool CGM}\label{coolCGMprop}
Once the corona is defined, we model the cool circumgalactic gas, assuming that this medium is composed by a population of clouds moving through the halo of M31. In particular, these clouds are subject to the gravitational force and to the drag force acted by the hot gas:
\begin{ceqn}
\begin{equation}\label{eq:dragforce}
\dot{\varv}_{\rm{drag}}=-\frac{\pi r^2_{\rm{cl}}\rho_{\rm{cor}}\varv^2}{m_{\rm{cl}}}\ ,
\end{equation}
\end{ceqn}
where $\rho_{\rm{cor}}$ is the coronal mass density, $\varv$ is the relative velocity between the corona and the cloud, $m_{\rm{cl}}$ is the cloud mass and $r_{\rm{cl}}$ is the cloud radius, determined by the pressure balance with the hot medium (see \citetalias{afruni19,afruni20}), assuming that the cool gas is at a temperature of $2\times 10^4$ K and that the clouds are spherical. From equation~\eqref{eq:dragforce} we can see how the efficiency of the drag force depends then on the mass of the clouds, which is the first free parameter of our models. We explore models with the cloud mass varying from $10^5$ to $10^8\ M_{\odot}$ (see Table~\ref{tab:dynesty}). Note that this mass does not vary during the motion of the cloud (this aspect is further discussed in Section~\ref{gasfate}).\\ 
As already mentioned, a second free parameter in all our models is the metallicity $Z$ of the cool gas, which can be constrained combining the hydrogen column densities predicted by our models with the observational information of the total silicon column densities (see Section~\ref{Observations} and~\ref{likelihood}). 
We let the metallicity free to vary from $0.003\, Z_{\odot}$ to $Z_{\odot}$ (Table~\ref{tab:dynesty}).\\
In this work we explore two different scenarios for the origin and dynamics of the cool CGM: outflow and inflow. These models are both described by three additional free parameters, which differ in the two cases and which are explained in detail in Sections~\ref{outmod} and \ref{modInf}.
\subsubsection{Outflow models}\label{outmod}
In this scenario, we assume that the cool clouds have originated in the ISM of M31 and have acquired kinetic energy from the supernova explosions in the galactic disc. We follow the same approach as in \citetalias{afruni20}. In brief, the clouds are ejected from different radii across the disc according to a star formation rate density. As in \citetalias{afruni20}, our fiducial models adopt the profile of \cite{pezzulli15} (but see Section~\ref{discussion}). The initial velocity is the vector composition of (i) the circular velocity of the disc (see Figure~\ref{fig:Rotcurv}) and (ii) an ejection velocity, defined in a cone around a direction perpendicular to the disc.\\
In addition to the cloud mass and the gas metallicity, we introduce for the outflow model three other free parameters: the initial ejection velocity $\varv_{\rm{kick}}$, the aperture of the outflowing cone $\theta_{\rm{max}}$ (if $\theta_{\rm{max}}=90^{\circ}$ the outflows are isotropic) and the mass loading factor $\eta$. This last parameter represents the ratio between the mass outflow rate $\dot{M}_{\rm{out}}(t)$ and the star formation rate $\mathrm{SFR}(t)$ of the galaxy, $\eta=\dot{M}_{\rm{out}}(t)/\mathrm{SFR}(t)$. Compared to similar star-forming galaxies at low redshift, M31 has currently a rather low star formation rate ($\rm{SFR}(0)=0.35\ M_{\odot} \rm{yr}^{-1}$, see \citealt{rahmani16}), which is unlikely to power strong outflows to the external regions of the halo. However, there are indications that the SFR was higher in the past.  In this work, we adopt as fiducial the star formation history (SFH) derived by \cite{williams17}, estimated by fitting stellar evolution models to color-magnitude diagrams of the Panchromatic Hubble Andromeda Treasury survey of the disc of M31, and characterized by 2 bursts of star formation around 2 and 8 Gyr ago.\\ 
The dynamics (i.e the orbits) of the clouds is uniquely defined by the choice of the three parameters $(m_{\rm{cl}}, \varv_{\rm{kick}}, \theta_{\rm{max}})$. For each combination of the parameters, we integrate 30 different orbits\footnote{This number has been chosen in order to sufficiently sample the galactic halo, keeping a reasonably low computational cost.} for 6 Gyr and, depending on the starting radius and ejection direction, the clouds will either escape the virial radius or fall back onto the disc after a time $t_{\rm{orb}}$ at which we stop the integration. Depending on the time $t_{\rm{cl}}$ elapsed along the orbits, the number density of clouds at that position is determined by the mass outflow rate at the relevant ejection time and hence proportional to $\mathrm{SFR}(t_{\rm{cl}})$, where $t_{\rm{cl}}$ is the loockback time.
   \begin{figure*}
   \includegraphics[clip, trim={0.2cm 0 0cm 0.2cm}, width=0.49\linewidth]{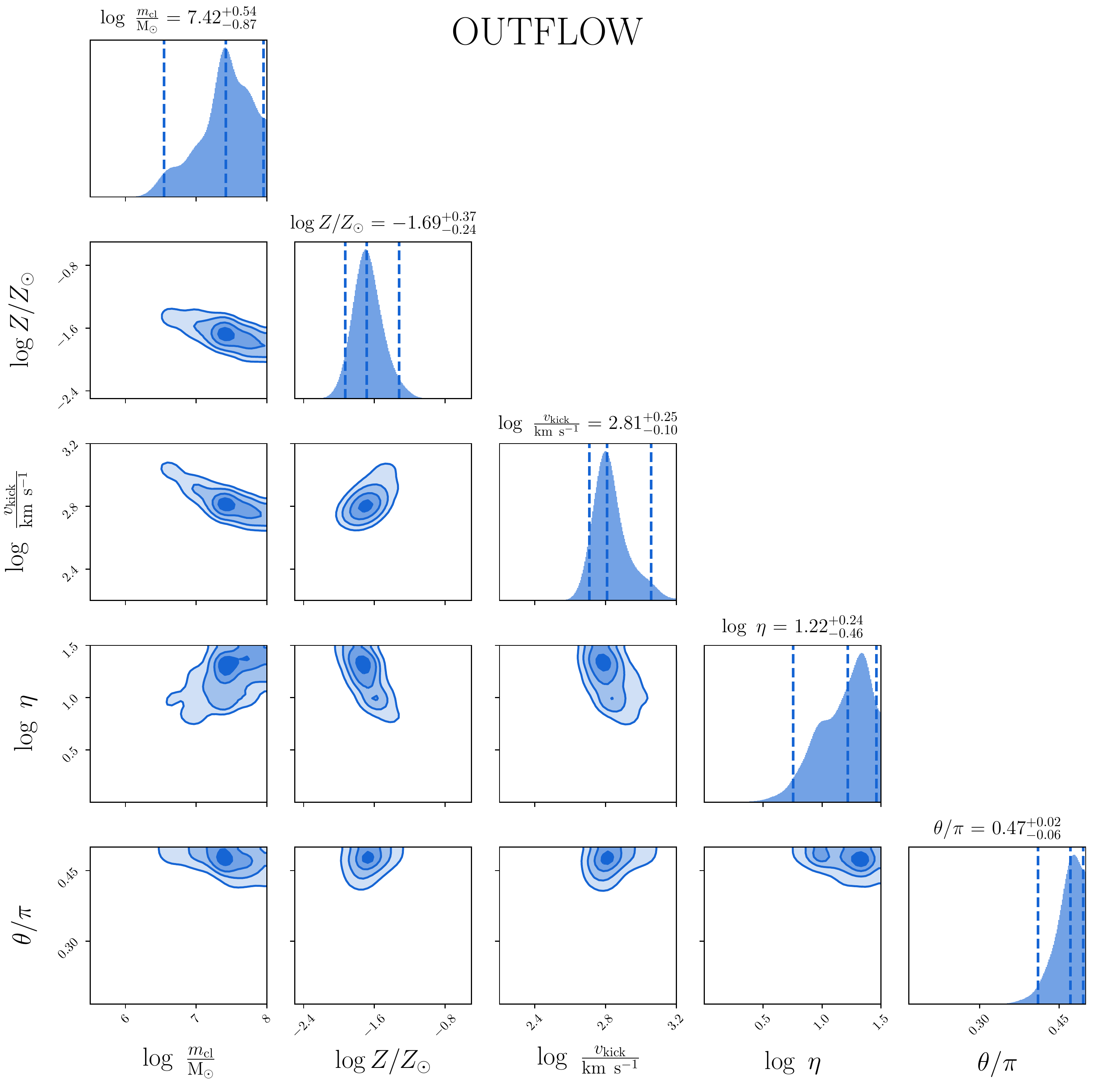}
   \includegraphics[clip, trim={0.2cm 0 0cm 0.2cm}, width=0.49\linewidth]{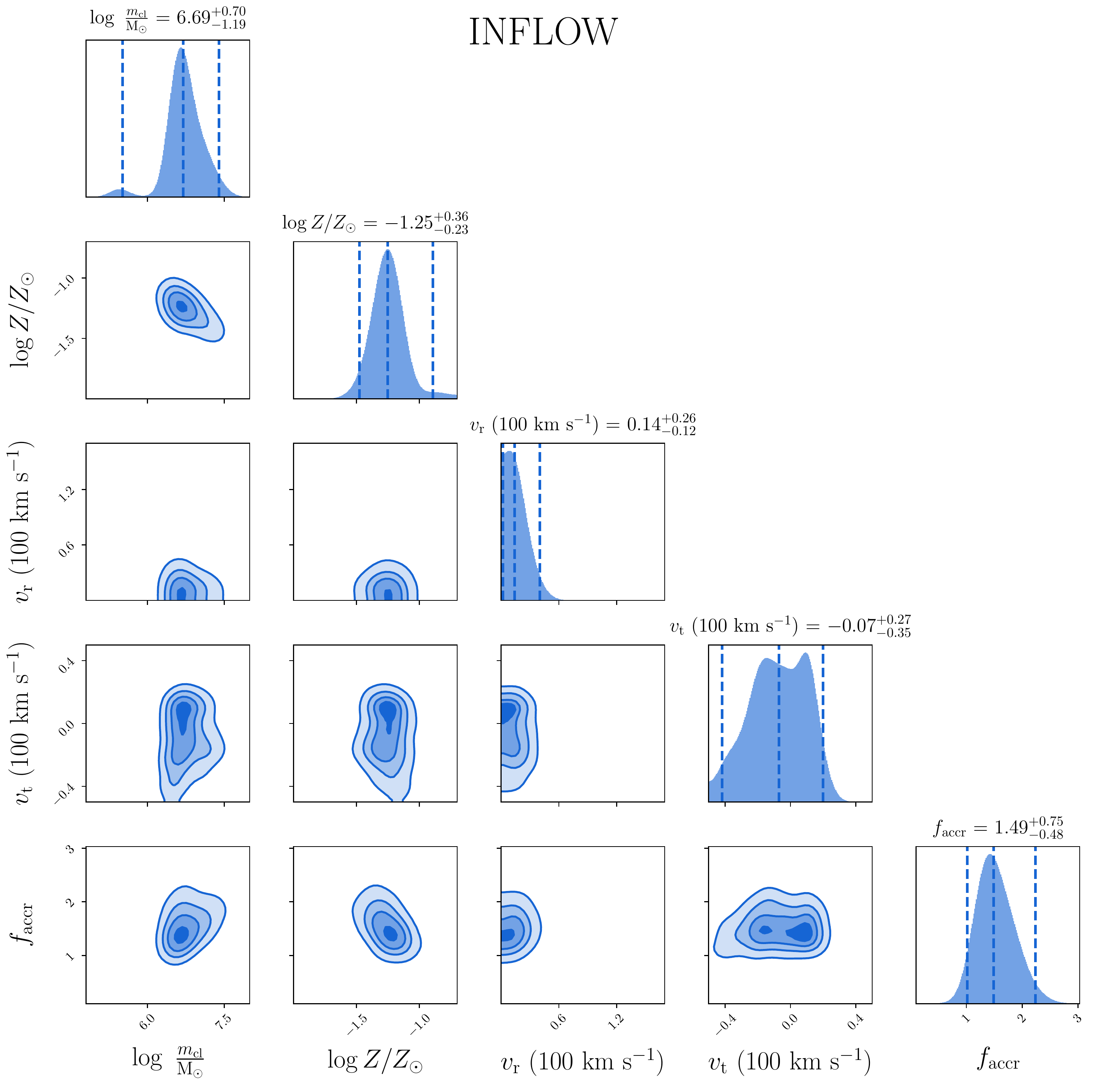}
   \caption{Posterior distributions from the Nested Sampling analysis, for the outflow scenario (left) and the inflow scenario (right), with reported the median values for each parameter distribution and the 2-sigma errors, which roughly correspond to the percentiles number 2.5 and 97.5.}
              \label{fig:Nested}%
    \end{figure*}
\subsubsection{Inflow models}\label{modInf}
In the second scenario that we explore in this work the CGM clouds are instead part of the cosmological accretion of gas onto the halo of M31 and are therefore inflowing from the external parts towards the galactic disc. In this case, we assume that the cool absorbers are starting from a spherical shell at the virial radius $r_{\rm{vir}}$ (see Section~\ref{Observations}) and are then infalling towards the central parts.\\
The three additional free parameters in this case are given by the initial radial and tangential velocity of the clouds ($\varv_{\rm{r}}$ and $\varv_{\rm{t}}$) and by the parameter $f_{\rm{accr}}=\dot{M}_{\rm{in}}/\dot{M}_{\rm{cosm}}$, where $\dot{M}_{\rm{in}}$ is the total cool gas mass rate of accretion into the halo and $\dot{M}_{\rm{cosm}}$ is the gas accretion expected from cosmological models. For the latter, we adopted the analytical prescription given in \cite{correa15a,correa15b}, which gives $\dot{M}_{\rm{cosm}}=11.5\ M_{\odot}\ \rm{yr}^{-1}$ at redshift $z=0$ for a halo with a mass $M_{\rm{vir}}$ equal to the one of M31 (see Table~\ref{tab:M31prop}) and accounting for the cosmological baryon fraction $f_{\rm{bar}}=0.158$ \citep[see][]{planck20}. The parameter $f_{\rm{accr}}$ therefore tells us how much the mass accretion rate of the cool CGM should differ from the cosmological predictions to reproduce the observations. We let this parameter free to vary, imposing however a gaussian prior centered on one (see Table~\ref{tab:dynesty}).\\
Also for this scenario we integrate, for each of the combination of the free parameters, 30 different orbits, each of them starting at a different polar angle from $r_{\rm{vir}}$, stopping the integration as soon as the clouds reach the central disc.
\subsection{Comparison with the observations}\label{likelihood}
To compare the predictions of our models with the data of the AMIGA project, we created synthetic observations for both the scenarios described above. In summary, we populate the orbits with a number of clouds given at each time by the cool gas mass outflow/inflow rate divided by the cloud mass. We then trace lines of sight in our model halo, at locations that correspond to the selected QSO sightlines (see Figure~\ref{fig:Obs}), and we calculate the line-of-sight velocities of the intersected model clouds (see \citetalias{afruni20} for more details). In order to be consistent with the observations, we select only clouds with line-of-sight velocities in the same range constrained by the AMIGA data ($\varv_{\rm{LSR}}$ ranging from -700 to -150 km $\rm{s}^{-1}$), excluding all the absorbers whose velocities could overlap with the Milky Way ($\varv_{\rm{LSR}}> -150\ \rm{km}\ \rm{s}^{-1}$) or the Magellanic stream, as described in Section~\ref{Observations}.\\ 
We also calculate the column density of each 'observed' model cloud, given by the formula 
\begin{ceqn}
\begin{equation}\label{eq:nsi}
N_{\rm{cl}}=2 n_{\rm{cl}} r_{\rm{cl}} \sqrt{1-(d_{\rm{cl}}/r_{\rm{cl}})^2}\ ,
\end{equation}
\end{ceqn}
where $r_{\rm{cl}}$ and $n_{\rm{cl}}$ are respectively the cloud radius and total number density, assumed constant throughout the cloud, and $d_{\rm{cl}}$ is the projected distance of the centre of the cloud from the intercepting sightline.\\
The last step needed to perform our analysis is the creation of a likelihood based on the comparison of synthetic and real observations. In order to do so we created, for each line of sight, a velocity distribution with the detected line-of-sight velocities (see \citetalias{afruni20} for more details), dividing the total velocity range in the same intervals used by \cite{lehner20p} for their AOD analysis (see in particular their Table 2). The likelihood is then composed of three different parts, dealing with respectively the kinematics ($\mathcal{L}_{\rm{kin}}$), the number of components ($\mathcal{L}_{\rm{num}}$) and the total silicon column densities ($\mathcal{L}_{\rm{dens}}$). 
The first term represents the Bayesian probability of each observed kinematic component given our model, while the second is obtained using the Poisson statistics to compare the number of components found in the observations and in our model for each line of sight. 
The third term in the likelihood quantifies the comparison, for each component, of the observed total silicon column density with the one predicted by our model. The latter is given by the sum of the column densities ($N_{\rm{cl,i}}$) of all the model clouds that have line-of-sight velocities in the selected velocity bin (which would appear as one single component in the observations), converted into a silicon column density by
\begin{ceqn}
\begin{equation}\label{eq:nsi}
N_{\rm{Si}}=\frac{1}{2.3}\left(\frac{Z}{Z_{\odot}}\right)\left(\frac{\rm{Si}}{\rm{H}}\right)_{\odot}\sum\limits_{i=1}^n N_{\rm{cl,i}} ,
\end{equation}
\end{ceqn}
where $n$ is the number of clouds in the selected velocity bin, $\log (\rm{Si}/\rm{H})_{\odot}=-4.49$ is the silicon solar abundance \citep[from][]{asplund09}, $(Z/Z_{\odot})$ is the gas metallicity and the factor 2.3 is used to transform the total numeric density into a hydrogen numeric density, assuming a fully ionized gas and the helium abundance reported in \cite{cox2000}.
For each line of sight then we defined $\mathcal{L}_{\rm{dens}}$ as:
\begin{ceqn}
\begin{equation}\label{eq:likedens}
\ln \mathcal{L}_{\rm{dens}}=-\frac{1}{2(n_{\rm{obs}})}\sum \frac{\left(\log{N_{\rm{Si}_{\rm{obs},i}}}-\log{N_{\rm{Si}_{\rm{mod},i}}}\right)^2}{\sigma^2}\\ ,
\end{equation}
\end{ceqn}
where $n_{\rm{obs}}$ is the number of observed components and $\sigma=0.15\ \rm{dex}$ is the uncertainty in the logarithm of the silicon column densities. This value represents a conservative choice with respect to the errors reported in \cite{lehner20p}, which have an average value of $\approx0.07\ \rm{dex}$, in order to take into account for SiIII lines close to saturation and for the presence of non-detections in the SiII and SiIV lines. Note that we perform this comparison only for the components that are classified as detections (in SiIII) by \cite{lehner20p}, as the treatment of non-detections is already included in $\mathcal{L}_{\rm{num}}$.\\
The final likelihood is given by the product of all the likelihoods calculated for each of the 23 lines of sight in our sample. We performed, for both the inflow and outflow scenarios, a Bayesian analysis using this likelihood and flat priors for all the parameters (except for a gaussian prior for $f_{\rm{accr}}$), exploring the parameter space in the ranges reported in Table~\ref{tab:dynesty}. In particular, we used the nested sampling method \citep[see][]{skilling04}, adopting the python package Dynesty \citep{speagle20}.
Like the more common Markov Chain Monte Carlo (MCMC) analysis, the nested sampling allows us to estimate the posterior distribution of our model outputs given the data, with the advantage of being more efficient in sampling multimodal distributions. With this analysis we were able to find the best choice of parameters that reproduces the AMIGA data, which we present in the following section.
\section{Results}\label{results}
In this section we outline the results and implications that we have obtained through the comparison of our models with the data of the AMIGA project, using a Bayesian analysis as described in Section~\ref{likelihood}. This allowed us to find for both the outflow and inflow scenarios the best choice of parameters needed to reproduce the observations.
\subsection{Outflow scenario}\label{resultsOut}
{ 
\renewcommand{\arraystretch}{1.5}
 \begin{table}
\begin{center}
\begin{tabular}{*{4}{c}}
\hline  
\hline
 Model & Parameter  & Prior (min, max) & Best-fit values\\
\hline 
\multirow{5}{*}{Outflow}&$\log m_{\rm{cl}}/M_{\odot}$
&5, 8&7.42\raisebox{.2ex}{$\substack{+0.54 \\ -0.87}$}\\
&$\log Z/Z_{\odot}$&-2.5, 0&-1.69\raisebox{.2ex}{$\substack{+0.37 \\ -0.24}$} \\
&$\log (\varv_{\rm{kick}}/\rm{km}\ \rm{s}^{-1})$&2, 4&2.81\raisebox{.2ex}{$\substack{+0.25 \\ -0.10}$} \\
&$\log \eta$&-1, 1.3&1.22\raisebox{.2ex}{$\substack{+0.24 \\ -0.46}$} \\
&$\theta/\pi $& $1/6$, $ 1/2$ &0.47\raisebox{.2ex}{$\substack{+0.02 \\ -0.06}$} \\
\hline
\multirow{6}{*}{Inflow}&$\log m_{\rm{cl}}/M_{\odot}$&5, 8& 6.69\raisebox{.2ex}{$\substack{+0.70 \\ -1.19}$}\\
&$\log Z/Z_{\odot}$&-2.5, 0& -1.25\raisebox{.2ex}{$\substack{+0.36 \\ -0.23}$}\\
&$\varv_{\rm{r}}\ (100\ \rm{km}\ \rm{s}^{-1})$&0, 5& 0.14\raisebox{.2ex}{$\substack{+0.26 \\ -0.12}$}\\
&$\varv_{\rm{t}}\ (100\ \rm{km}\ \rm{s}^{-1})$&-1, 1& -0.07\raisebox{.2ex}{$\substack{+0.27 \\ -0.35}$}\\
&$f_{\rm{accr}}$&0.1, 4& 1.49\raisebox{.2ex}{$\substack{+0.75 \\ -0.48}$}\\
\hline
\end{tabular}
\end{center}
\captionsetup{justification=centering}
\caption[]{Results of the nested sampling analysis for the two scenarios investigated in this work. All the priors are flat except for $f_{\rm{accr}}$, given by a truncated normal profile centered in 1 and with a width of 0.5. The last column shows the values of the 50th percentiles (median) of each parameter distribution, with the 2-sigma errors, which roughly correspond to the percentiles number 2.5 and 97.5.}\label{tab:dynesty}
\end{table}
}
   \begin{figure*}
   \includegraphics[clip, trim={0.2cm 0 0cm 0.2cm}, width=0.49\linewidth]{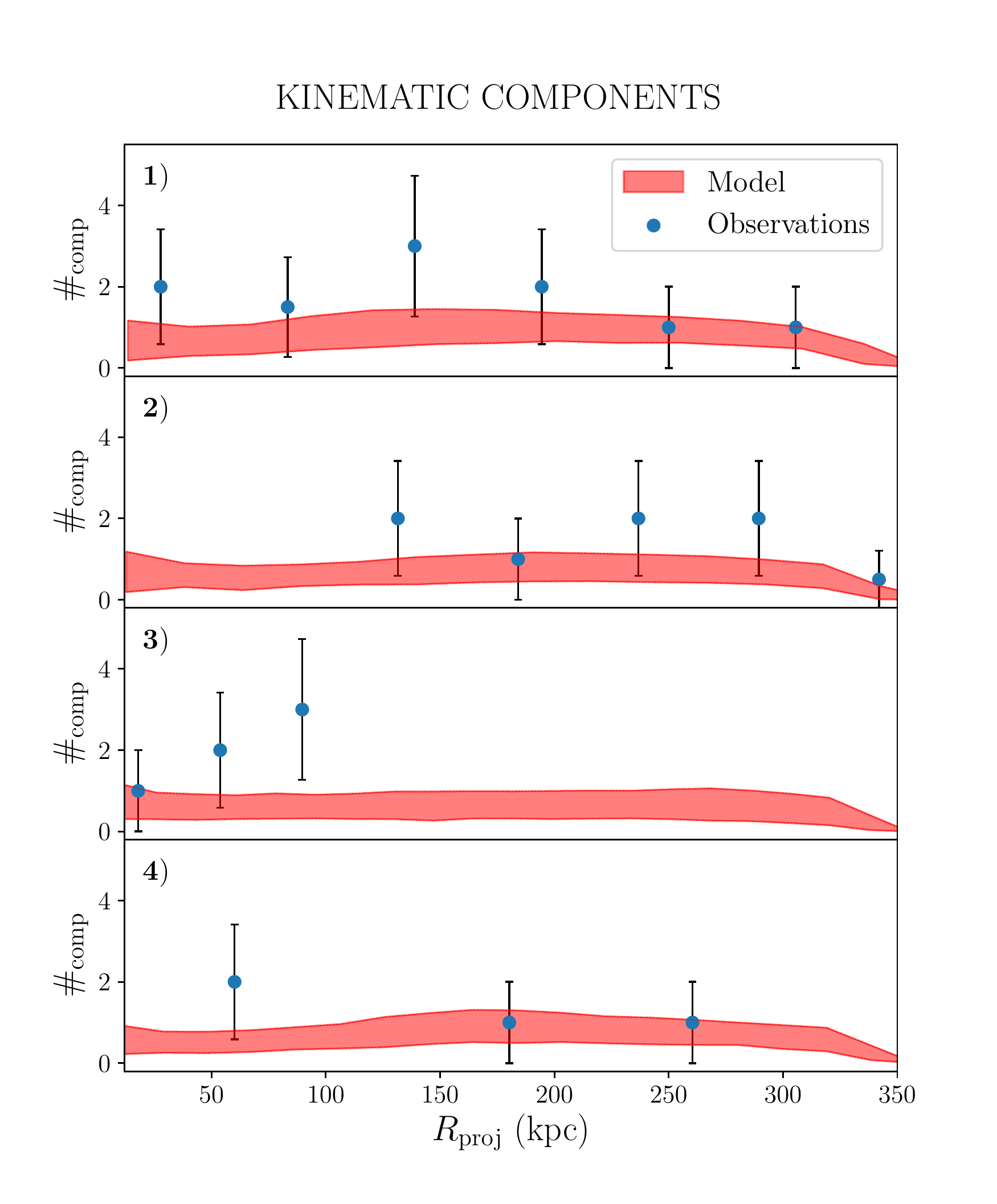}
   \includegraphics[clip, trim={0.2cm 0 0cm 0.2cm}, width=0.49\linewidth]{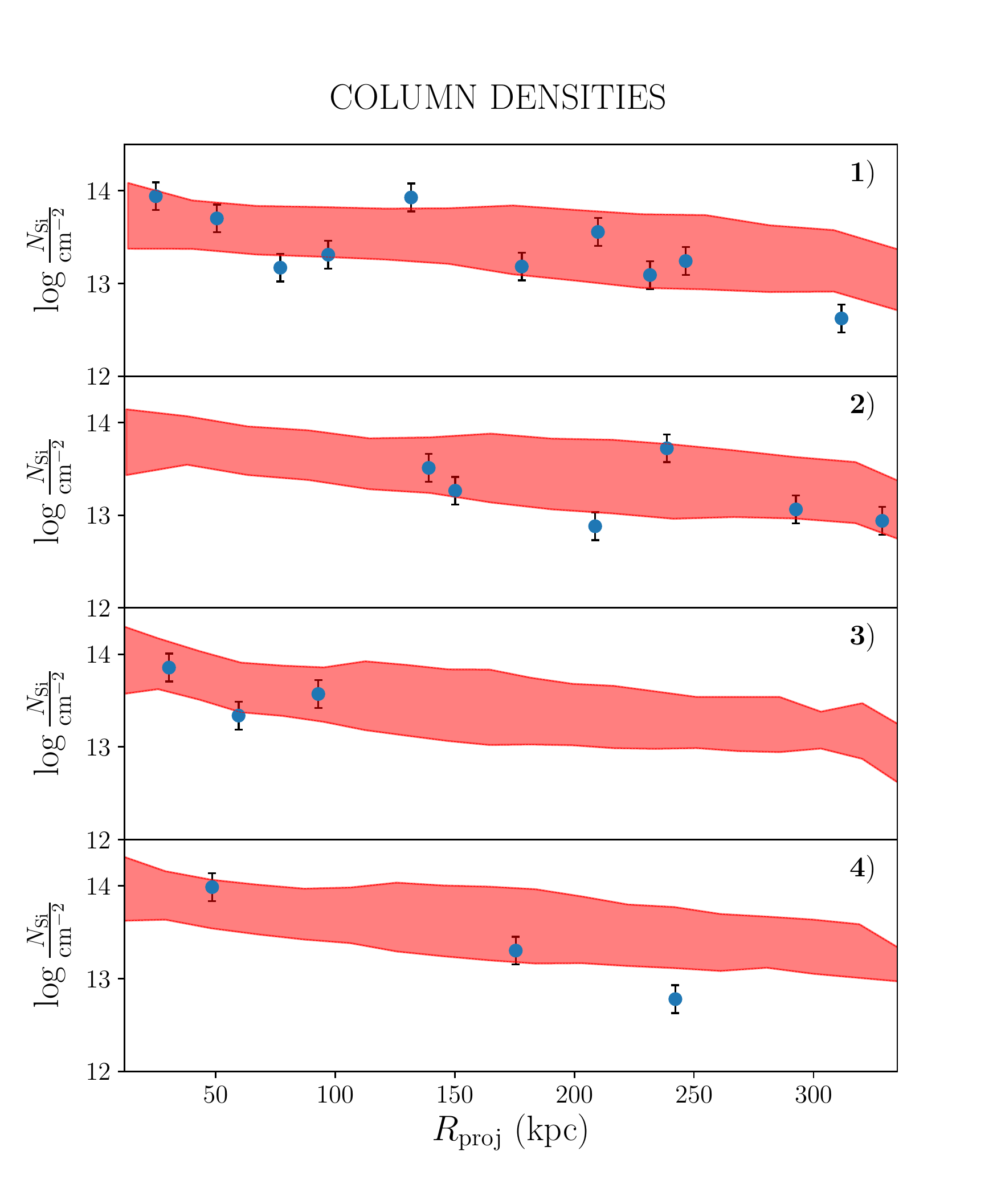}
   \caption{Comparison between the predictions of the best-fit models described in Section~\ref{resultsIn} (red bands) and the observations (blue points), with the number of components on the left and the total silicon column density on the right, both as a function of the projected distance from M31. We show, going from top to bottom, the results for the four quadrants outlined in Figure~\ref{fig:Obs}, from 1 to 4. The observed number of components is binned over the projected distance in bins of about 55 kpc, a value chosen in order to have a good enough statistics in the data. The width of the bands represents the standard deviation calculated for the different predictions of the 50 models used for the comparison. The error bars in the number of components represent the poissonian error, while for the column densities they are equal to 0.15 dex.}
              \label{fig:Comparison}%
    \end{figure*}
   \begin{figure*}
   \includegraphics[clip, trim={0.2cm 0 0cm 0.2cm}, width=\linewidth]{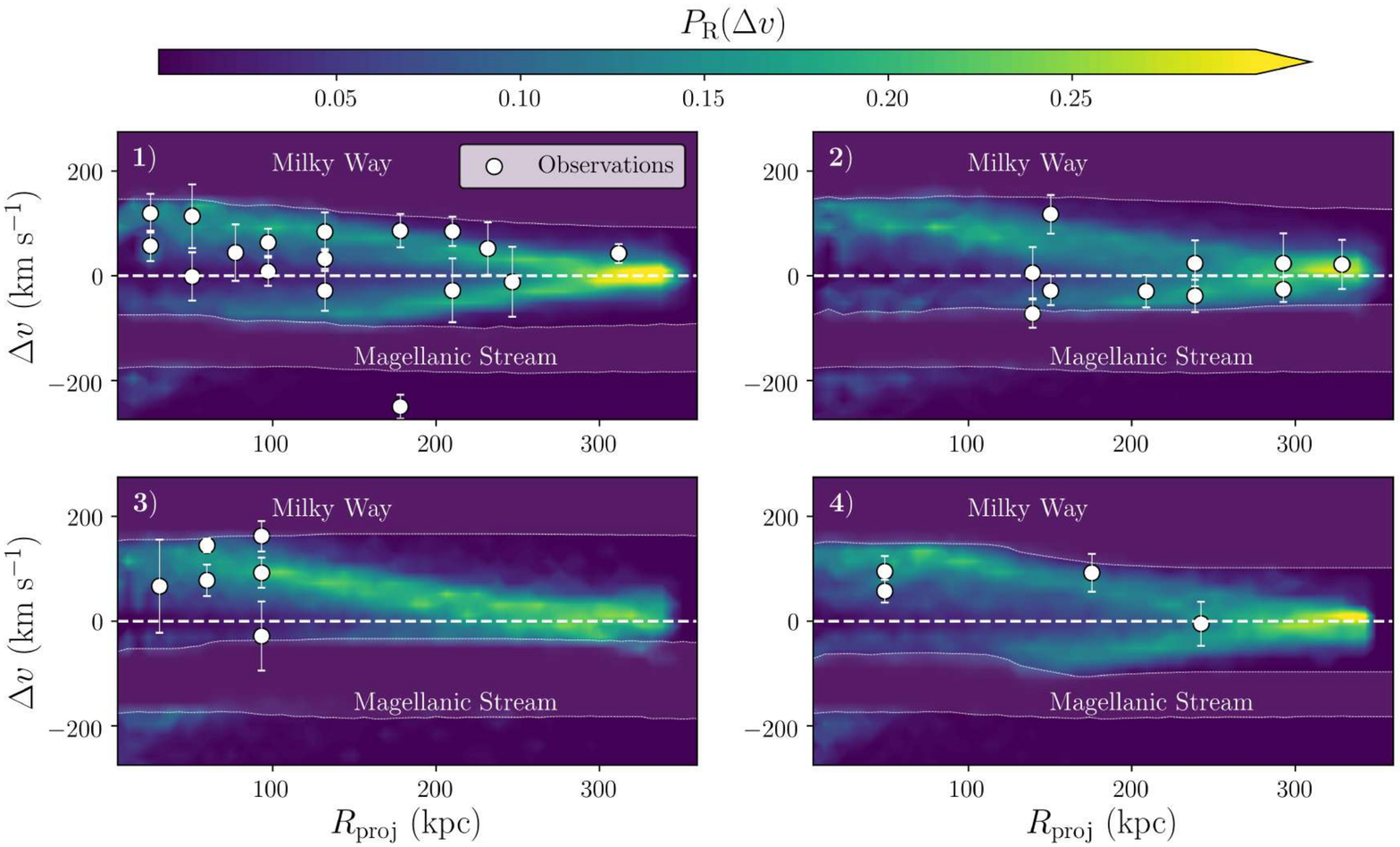}
   \caption{Comparison between the kinematics predicted by the best-fit inflow models described in Section~\ref{resultsIn} (colormap) and the observed kinematic components (white points), as a function of the projected distance from M31, for the four quadrants outlined in Figure~\ref{fig:Obs}. The error bars on the data points show the velocity boundaries of each absorption component, as identified by \protect\cite{lehner20p}. The model predictions are calculated as explained in Section~\ref{resultsIn} and represent, for each projected radius, the normalized probability distribution $P_{\rm{R}}(\Delta \varv)$ to observe a cloud at a given line of sight velocity (with respect to the systemic velocity of M31). We also indicate in each panel the velocity ranges affected by contamination from the Magellanic Stream and the Milky Way, excluded from both observations and models.}
              \label{fig:Comparison1}%
    \end{figure*}
The results for the outflow model are reported in the left-hand panel of Figure~\ref{fig:Nested} and in Table~\ref{tab:dynesty}. In order to reproduce the observations of the cool circumgalactic gas, the outflowing clouds need to be very massive, with $m_{\rm{cl}}\gtrsim10^7 M_{\odot}$. 
In addition to the high cloud mass, we find that the global galactic winds need to be nearly isotropic ($\theta_{\rm{max}}\sim 90^{\circ}$) and that, most importantly, both the mass loading factor $\eta$ and the initial ejection velocities $\varv_{\rm{kick}}$ require large values in order to reproduce the observations.\\
Following \citetalias{afruni20}, we can use these values to calculate the total kinetic power needed by the wind, given by:
\begin{ceqn}
\begin{equation}\label{eq:outpower}
\dot{K}_{\rm{out}}= \frac{1}{2} \dot{M}_{\rm{out}}\ \varv_{\rm{kick}}^2\ ,
\end{equation}
\end{ceqn}
where $\dot{M}_{\rm{out}}=\eta \mathrm{SFR}(t)$. This can be compared to the kinetic energy available per unit time from the supernova explosions in the disc \citep{cimatti19}, in order to obtain the efficiency $f_{\rm{SN}}$ of the supernovae in transferring kinetic energy to the wind (see \citetalias{afruni20}).
Through this comparison, we found that to power the winds predicted by our models we need efficiencies of 730\%. This is clearly unphysical, since they imply that the outflows would need significantly more energy than the one available from the supernovae. On the contrary, as most of the supernova energy is expected to be radiated \cite[e.g.][]{mckee77,kim15}, the theoretical expectation for $f_{\rm{SN}}$ is of the order of 10\% \citep[e.g.][]{martizzi16}, or at most a few 10\% in the case of clustered supernova explosions \citep[][]{fielding18}. Therefore we conclude that SN driven galactic outflows are not a viable way to reproduce the majority of the cool gas observed in the halo of M31.  This is in line with the main result of \citetalias{afruni20}, where we found that galactic winds cannot reproduce the cool CGM of a sample of about 40 low-redshift star-forming galaxies. Note that in our previous study only one line-of-sight per galaxy was available, while for M31 there are more than 20 QSO sighlines within its virial radius (in projection). Moreover, in our current model we also include the effect of a burst of star formation 2 Gyr ago, observed in the disc of M31 \citep[][]{williams17} as explained in Section~\ref{outmod}. This however does not change our main result that star-formation-driven outflows cannot reproduce the majority of the cool CGM.\\
As a final note, we also find that the metallicity required by this model is extremely low ($Z<0.05\,Z_{\odot}$, as opposed to $Z\sim Z_{\odot}$ as expected for gas originated in the ISM), which is pointing to a further inconsistency with what we would expect for gas ejected from the central galaxy. We therefore exclude the outflow scenario as the main source of the cool CGM of M31 and we conclude that (at least the vast majority of) this medium has a different origin, which we will explore more in detail in the following section. We do not exclude however the possibility that the outflows might affect the CGM in the internal regions of the halo, close to the galactic disc, as we discuss in Section~\ref{outDiscuss}.
\subsection{Inflow scenario}\label{resultsIn}
On the right-hand panel of Figure~\ref{fig:Nested} and in Table~\ref{tab:dynesty} we show the results of the nested sampling analysis for the inflow models, whose features are described in Section~\ref{modInf}. We can see that there is a very well defined region of the parameter space that allows us to reproduce the AMIGA observations. In particular, we find that the data are well described by low-metallicity ($Z\approx0.05\,Z_{\odot}$) clouds with a mass of about $5\times10^6\ M_{\odot}$, accreted at a rate only slightly higher than the cosmological gas accretion rate expected at $z=0$ ($f_{\rm{accr}}\approx1.5$) and with low tangential and radial velocities as they start their infall at the virial radius. We go more in the details of the physical picture arising from this parameter choice in Section~\ref{infDynamics}.
\subsubsection{Comparison with observations}
The comparison between data and model predictions is shown in Figures~\ref{fig:Comparison} and \ref{fig:Comparison1}. Our model of the circumgalactic medium of M31 is not isotropic, since the hot gas is rotating (see Section~\ref{hotgasMod} and Appendix~\ref{hotgas}) and the cool clouds are not infalling perfectly radially towards the galaxy. We therefore show our results separately for the four quadrants defined in Figure~\ref{fig:Obs}, in order to investigate any azimuthal dependence in the model and in the comparison of our outputs with the observational data. 
In the left- and right- hand panels of Figure~\ref{fig:Comparison} we show respectively the number of kinematic components and the total silicon column density (see Section~\ref{Observations}), both as a function of the projected distance from M31. The blue points represent the observations, while the red bands represent the predictions of our models. To obtain these predictions, we averaged the results of 50 models with values for the free parameters sampled from the posterior distributions shown in the corner plot of Figure~\ref{fig:Nested} (right-hand panel). We can see how in general the observations are well reproduced by our models for all the four quadrants of the observational plane and there seems not to be any strong azimuthal dependence in the properties of the cool gas, both for model and data.
The number of components is however slightly underpredicted by our models. Models with significantly higher accretion rates formally perform better in this respect, but are discarded by our prior on the cosmological accretion.\\
There are a couple of assumptions of our model that can have an impact in the comparison between measured and predicted number of components (see also Section~\ref{origins}).
First, our clouds are assumed to remain spherical and in pressure equilibrium with the corona throughout their infall. This implies that in the inner regions the clouds are smaller (lower cross section) than in the outer halo. Hydrodynamical simulations \citep[e.g.][]{armillotta17} indicate, instead, that clouds tend to quickly lose their spherical shape and to be significantly more extended than what we assume, being therefore easier to intercept by a line of sight from some viewing directions. The second assumption of our model that is worth mentioning in this context is the isotropy of the accretion. Because we are assuming that the accretion is isotropic, the number of components predicted by our models should be better understood as an average over spherical shells. If the accretion were anisotropic, the number of components in a certain region could be higher than this average value. Unfortunately, it is not easy to directly test this possibility with the available data, as we cannot probe those directions that are not intercepted by a quasar sightline, or which are subject to contamination from the Milky Way (see Figure~\ref{fig:Obs}).\\
In the four panels of Figure~\ref{fig:Comparison1} we focus on the comparison of the kinematics (line-of-sight velocity of the cool clouds with respect to the systemic velocity of M31), as a function of the projected radius, predicted by our models and observed in AMIGA. In addition, we show the velocity ranges in these position-velocity plots that are affected by the presence of the Milky Way and the Magellanic Stream. These ranges have been excluded both from the observations and in our models. The observations are shown as white points and correspond to the kinematic components reported in the upper panel of Figure~\ref{fig:Obs}, with the error bars denoting the velocity ranges associated to each identified absorbing system \citep[see][]{lehner20p}, also corresponding to the velocity bins that we used to create our likelihood (see Section~\ref{likelihood}). In the colormap we report instead the prediction for the same 50 models used for Figure~\ref{fig:Comparison}, showing, for each line of sight, what is the probability that an intersected cloud has a certain line-of-sight velocity. In particular, the entire velocity range is divided in bins of $20\ \rm{km}\ \rm{s}^{-1}$ and the colormap shows the probability of a component to lie in one of these bins.\footnote{Note that all probabilities are normalized in order to have, for each line of sight (i.e. each vertical slice in the diagram), a total probability of 1.
The brightest colors in the colormaps are located at large projected radii because at these distances the range of the predicted velocities is smaller, not because in these regions it is more likely to intersect a cloud (see Figure~\ref{fig:Comparison} for the distribution of components as a function of projected distance).}
The model predictions match remarkably well, except for a few outliers, the velocities observed for the cool circumgalactic gas. In particular, they show the same pattern of the data of having, on average, the higher line-of-sight velocities at small projected radii. We note that a few observed components at $R_{\rm{proj}}<150\ \rm{kpc}$ tend to have absolute values of the velocities slightly lower than what preferred by our models. This could be due to second-order hydrodynamical effects that, by deforming and possibly disrupting the clouds (see also Section~\ref{gasfate}), especially in the last part of their infall, would slow down the cool gas more efficiently than what predicted by our semi-analytical models. We will study more in detail these interactions in a forthcoming paper.\\
We have therefore found that there is a class of inflow models that are cosmologically feasible ($f_{\rm{accr}}\approx1.5$) and that can successfully describe the observational data. Interestingly, our models also predict (except for $R_{\rm{proj}}\lesssim50$ kpc) a lack of absorbers at velocities $\varv_{\rm{M31}}\lesssim-200\ \rm{km}\ \rm{s}^{-1}$, which are not affected by the Magellanic Stream and where also the AMIGA observations show a lack of detections (with only one exception).\\
Finally, a fundamental property that we are also able to constrain with the comparison between model predictions and observations is the metallicity of the cool gas, which sets the normalization of the silicon column density profile shown in the right panel of Figure~\ref{fig:Comparison}. The Bayesian analysis returns a metallicity lower than $0.1\ Z_{\odot}$, which is consistent with gas poorly enriched by star formation processes, as expected for material that is accreting from the intergalactic medium (see Section~\ref{Infdisc}). This is consistent with our assumption of cosmological accretion and hence confirms that this is the most likely origin of the majority of the cool CGM around M31. In Section~\ref{Infdisc} we compare our estimate with the one obtained by \cite{lehner20p}.
   \begin{figure}
   \includegraphics[clip, trim={0.3cm 0.1cm 1.5cm 1.5cm}, width=\linewidth]{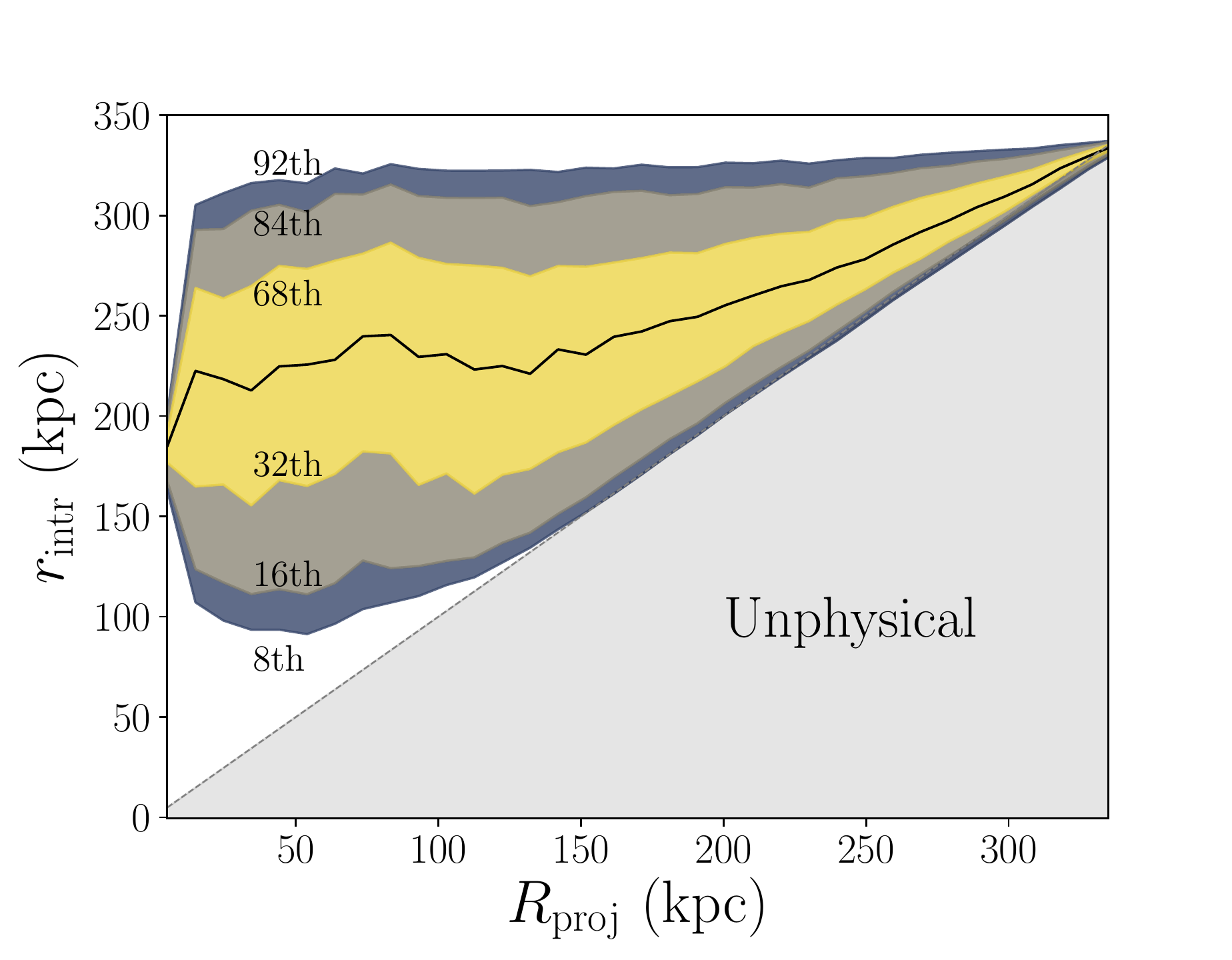}
   \caption{Distribution of the intrinsic galactocentric distances of the clouds as a function of their projected separation from M31, as predicted by our best-fit inflow model. The black curve shows the median values of the distributions of intrinsic distances of the model clouds ('observed' as explained in Section~\ref{likelihood}) at each projected radius, with the color bands showing different percentiles (8th, 16th, 32nd, 68th, 84th, 92nd). The region where $R_{\rm{proj}}>r_{\rm{intr}}$ is necessarily empty, as the projected distance is a physical lower limit of the intrinsic one. Note how the majority of the clouds, even at small projected distances, are located at intrinsic distances larger than 150 kpc, suggesting that the cool absorbers are mainly located in the external parts of the halo of M31.}              \label{fig:CloudDist}%
    \end{figure}
   \begin{figure}
   \includegraphics[clip, trim={0.2cm 0 0.5cm 1cm}, width=\linewidth]{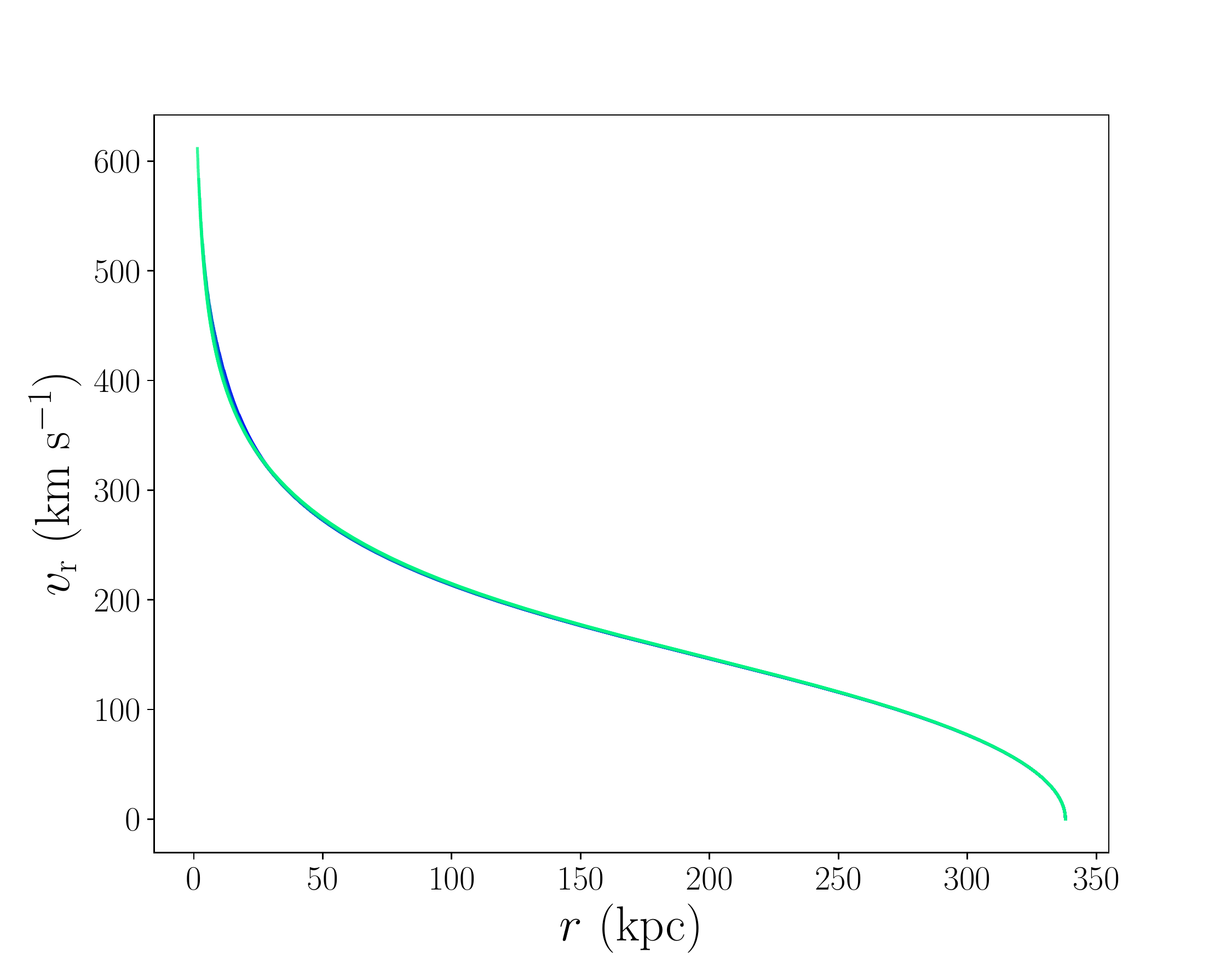}
   \includegraphics[clip, trim={0.2cm 0 0.5cm 1cm}, width=\linewidth]{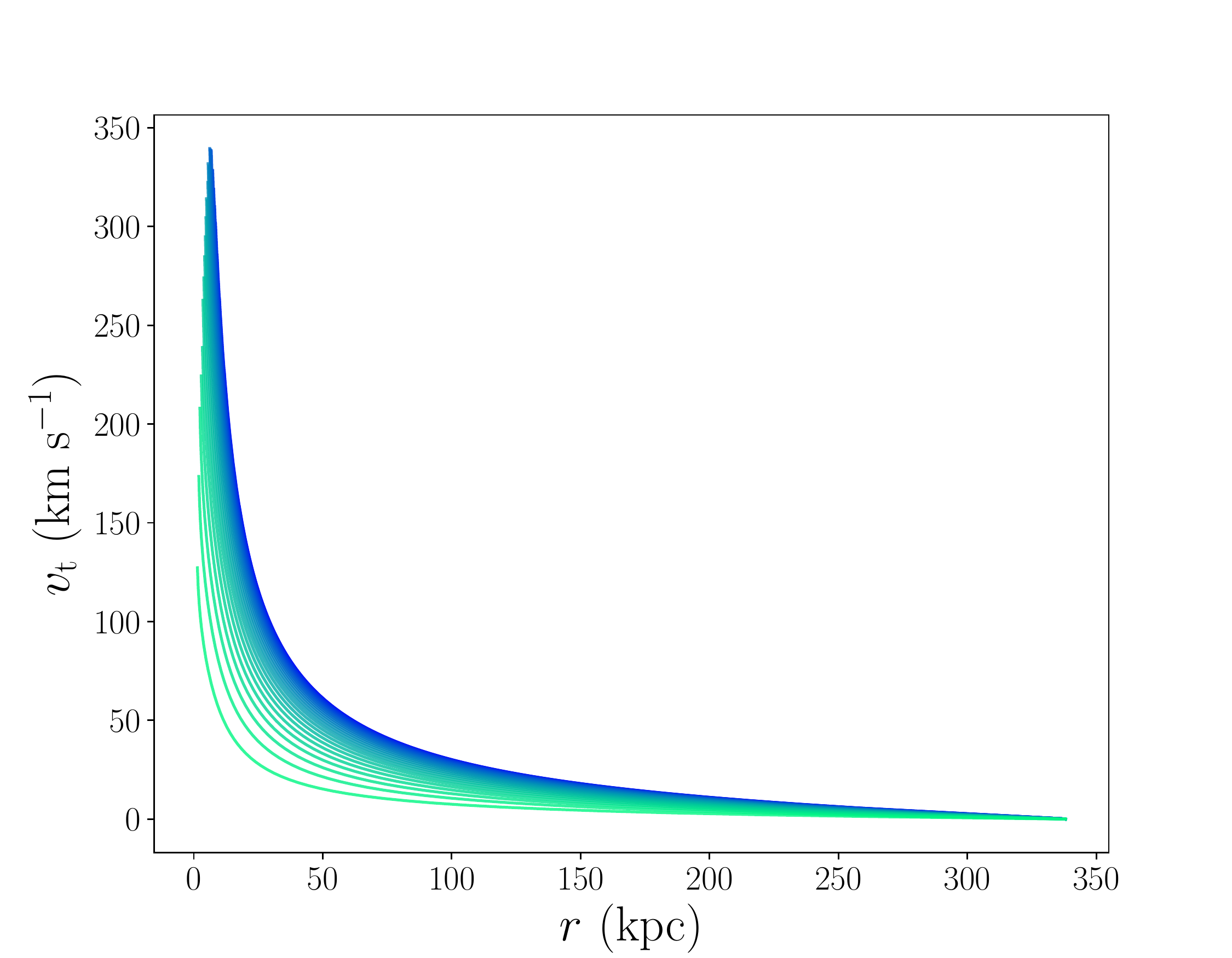}
   \includegraphics[clip, trim={0.2cm 0 0.5cm 1cm}, width=\linewidth]{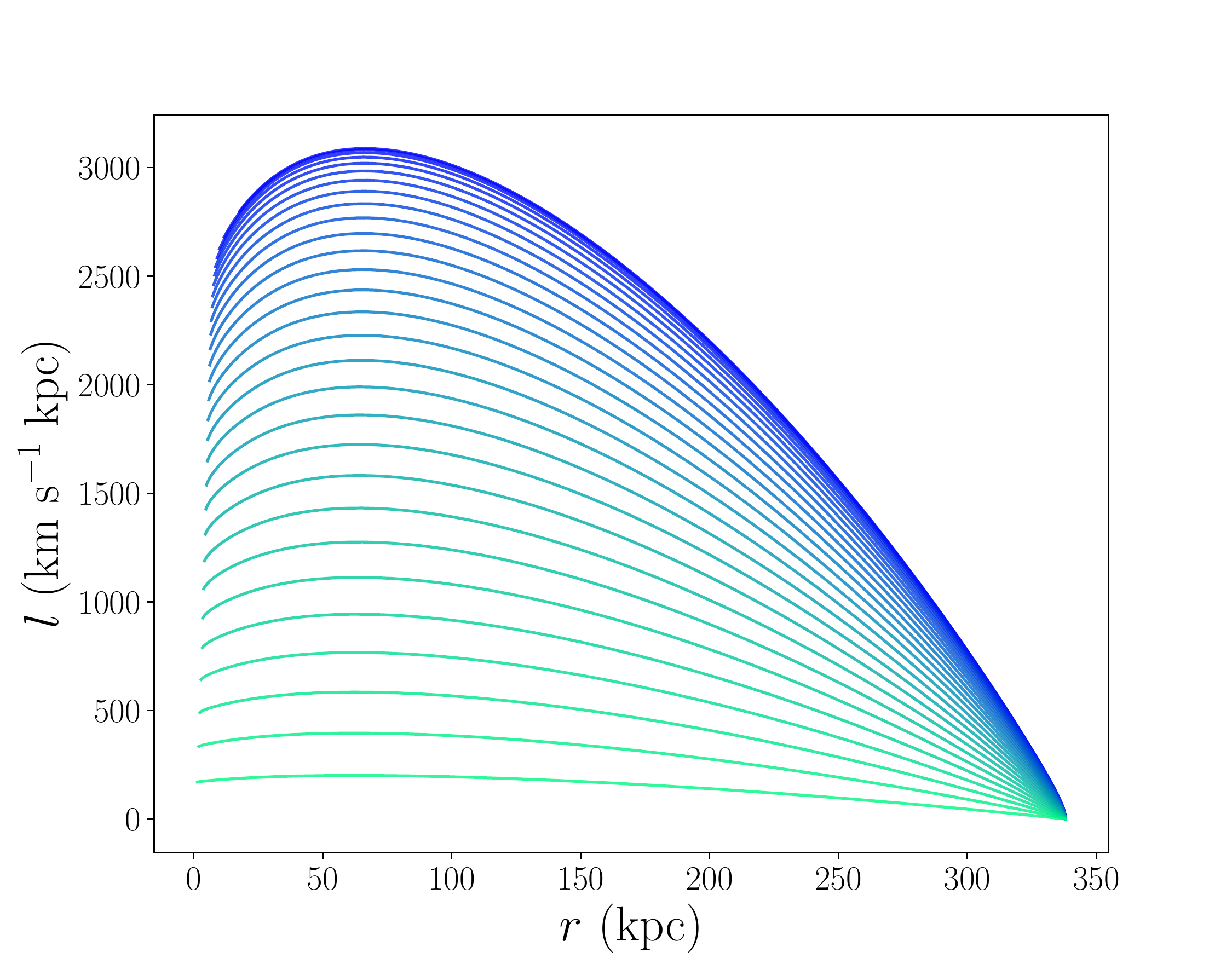}
   \caption{Radial velocity (top), tangential velocity (centre) and specific angular momentum (bottom) of the clouds, as a function of the intrinsic distance from the centre and as predicted by an inflow model with a choice of parameters representative of the best-fit parameter region reported in Table~\ref{tab:dynesty}. We show different orbits for clouds starting at different polar angles from a shell at the virial radius of M31. The colour of the lines changes with increasing initial cylindrical radius $R$ of each orbit, going from 0 (light green) to $r_{\rm{vir}}$ (dark blue).}
              \label{fig:Tangential}%
    \end{figure}\\
\subsubsection{Intrinsic and projected distances of the absorbers}
In Figure~\ref{fig:CloudDist} 
we show the distribution of the intrinsic distances of the intersected clouds as a function of their projected distances from the centre. This is in particular the prediction of the same 50 best-fit models used to create Figures~\ref{fig:Comparison} and \ref{fig:Comparison1} and considering all the azimuthal angles. More in detail, the black curve represents the median of this distribution, with the colours that show instead different percentiles as indicated. Obviously, the projected distance is only a lower limit of the intrinsic one, as depicted by the gray shaded area. More interestingly, from this distribution it is evident how  the galactocentric distance of a cloud can be much larger than the one that is seen in projection. In particular, note how the vast majority of the clouds that we observe are located beyond 150 kpc from the disc of M31.\\
This finding is caused by two effects: (i) the cloud size increases with the distance from the centre, since in the external regions the clouds are in pressure equilibrium with a lower density hot gas (see Figure~\ref{fig:Corona}) and it is therefore more likely to intercept these absorbers, given their higher cross-section; (ii) as we will see below, the cool CGM is infalling at lower velocities in the external regions of the halo, therefore the orbits are populated with more clouds at these distances.\\
This result clearly shows how, in absorption studies of the CGM, the projected locations of the lines of sight might be not representative of the real distances of the absorbing gas systems from the central galaxy. We discuss more in details the implications of this finding in Section~\ref{gasfate}. 
\subsubsection{Dynamics and properties of the cool clouds}\label{infDynamics}
We now discuss in some greater detail the physical properties and the dynamical behaviour of the cool CGM absorbers, according to the predictions of our best-fit inflow model. The infalling clouds have a mass of about $5\times10^6\ M_{\odot}$ and an initial radius $r_{\rm{cl}}\simeq5.7$ kpc. Under our assumption that they have a constant mass with time (but see Section~\ref{gasfate}) and are pressure confined by the hot corona, these clouds tend to shrink while falling towards the disc, reaching sizes smaller than 1 kpc in the central ($r\lesssim20$ kpc) regions. We have also performed a test where the cloud masses at the virial radius are extracted from a power-law distribution, whose slope, minimum and maximum cloud mass were left free to vary as free parameters. In this case we find that the slope of the power law is unconstrained and that the range of possible cloud masses is small (0.7 dex) and centered on the best-fit mass found with our fiducial model. Therefore, allowing the clouds to have a more complex mass distribution does not seem to affect our results.\\ 
The total mass infall rate of the cool gas is equal to $15.3\ M_{\odot} \rm{yr}^{-1}$ and the average time needed for the clouds to reach the disc from the virial radius is of about 2.5 Gyr. Since the mass accretion is constant with time in our models, the total amount of cool gas in the halo of M31 is therefore $\approx4\times10^{10}\ M_{\odot}$. Note that, as we have seen in Section~\ref{hotgasMod}, the sum of the corona and the stellar component of M31 accounts for almost $60\%$ of the total expected baryons. Therefore, since we predict that $\approx15\%$ of the baryons are in the cool CGM component, in our model the total baryonic mass within the halo of M31 is very close to the expected cosmological fraction.\\
The three panels of Figure~\ref{fig:Tangential} show the overall intrinsic dynamics of the clouds in our best-fit models, for 30 orbits described by our best-fit parameters, all starting at the virial radius but from different positions across the virial sphere. On the top panel, we report the cloud radial velocities as a function of the galactocentric distance: in this case, the profiles for the 30 different orbits overlap almost perfectly with each other, reflecting that the anisotropies present in our model have only a minor impact on the radial motion of the clouds. At the virial radius the clouds start with a very low velocity: for this plot we adopted $\varv_{\rm{r}}=14\ \rm{km}\ \rm{s}^{-1}$, which is the median value from the posterior distribution obtained with the nested sampling. In order to reproduce the observational data, the initial velocity needs indeed to be lower than a few tens of km $\rm{s}^{-1}$ (see table~\ref{tab:dynesty}).
The clouds then accelerate during their infall, maintaining however a relatively low speed due to the drag force acted by the corona and reaching high velocities only in the central regions, where the gravitational force becomes stronger and the drag force weaker due to the compression of the clouds leading to a smaller cross section.\\
The clouds described by our best-fit models are not falling only radially towards the galactic disc as their motion has a tangential component. From the nested sampling analysis we found a median value for the initial tangential velocity of $-7\ \rm{km}\ \rm{s}^{-1}$, with uncertainties of about $30\ \rm{km}\ \rm{s}^{-1}$. Note that this range is consistent with the cool clouds having an initial angular momentum equal to or even larger than that of the corona, though we cannot determine whether this is the case with the available data. The profiles of the tangential velocities and of the specific angular momentum as a function of the intrinsic distance from the centre are shown respectively in the central and bottom panels of Figure~\ref{fig:Tangential}. These have been obtained assuming an initial $\varv_{\rm{t}}=0\ \rm{km}\ \rm{s}^{-1}$ (which is within the uncertainties of our posterior for the initial tangential velocity). The drag force progressively drives the clouds towards co-rotation with the hot CGM (therefore also on the same sense of rotation of the disc, see Section~\ref{M31setup} and \ref{hotgasMod}) with the two gas phases exchanging angular momentum. 
The cool medium tends to acquire angular momentum from the hot gas, reaching values of a few thousands of km s$^{-1}$ kpc, comparable with the angular momentum of the hot corona (see Figure~\ref{fig:angmomHot}).
After acquiring significant angular momentum from the outer parts of the corona, the cool gas continues its infall motion towards the innermost regions, where the hot gas is relatively poor in angular momentum, and eventually deposits back, at a different location, part of the angular momentum that was previously acquired (see bottom panel of Figure~\ref{fig:Tangential}). We note in passing that in this way cool accreting flows potentially provide a mechanism to transfer angular momentum from the outermost to the innermost regions of the hot halo (though further exploring the possible implications of this mechanism is beyond the scope of this work). After an infall of about 2.5 Gyr, our model predicts that the clouds cross the plane of the disc (defined as z = 0) at cylindrical radii between about 1 and 18 kpc. Note however that in reality clouds may not be able to reach such small galactocentric distances (see Section~\ref{gasfate}). 
\section{Discussion}\label{discussion}
In the previous section we have seen that our analytic method indicates a clear preference for an inflow origin, as opposed to an outflow origin, for the cool CGM of M31. It also provides clear predictions for some of the most important properties of the circumgalactic gas of M31. Here, we discuss the implications and limitations for both scenarios of outflow and inflow and we draw conclusions regarding the origin and fate of the cool CGM and its connection with the central galaxy.
\subsection{Outflow}\label{outDiscuss}
The first result of this study is the inability of a galactic wind to explain (the majority of) the cool CGM of M31, mainly because of the unphysical energy requirements that are needed in order to reproduce the data. It is important however to remark that the outflows considered in this work are represented by an idealized flow of gas clouds at a constant temperature of $2\times 10^4$ K. Observations \citep[e.g.][]{strickland07,martin12} and pc-resolution simulations of supernova-driven winds \citep[e.g.][]{li17,kim18,fielding18,kim20}, focused on a region that goes from the disc up to a few kpc above it, have shown instead how outflows seem to be multiphase, with most of the gas mass residing in the cool phase and most of the energy in the hot phase of the wind \citep{li20}.\\ The entrainement and acceleration of the clouds by fast-moving hot winds \citep[see][]{schneider20} could in principle help in bringing the cool gas up to the large galactocentric distances probed in this work. Moreover, this uplifting of the cool medium could happen also because of buoyancy of entropy-driven, slow-moving hot winds \citep[see][]{keller20}.
We could expect however that in order to be able to do so, the total energy (kinetic plus thermal) injected to the hot wind would need to be comparable to the one that we have found in Section~\ref{resultsOut}, to overcome the gravitational force and the pressure of the pre-existing corona. 
Therefore, even though the absence of this effect is a limitation of our models, we consider unlikely that it would strongly affect our main findings and change our conclusions. An accurate modeling of the multiphase winds, as well as a more proper treatment of non-thermal effects such as those associated to cosmic rays \citep[e.g.][]{pfrommer17,hopkins21}, 
is outside the scope of this work and we leave it for future studies.\\
Another possible source of uncertainty is in the choice of the star formation rate density profile. As in \citetalias{afruni20}, we adopted the profile predicted by \cite{pezzulli15} (see also Section~\ref{outmod}), which is however only an approximation of the one observed in the disc of M31. One feature that is not reproduced by this profile is the enhanced star formation rate in a ring roughly between 10 and 12 kpc \citep[e.g.][]{roblesValdez14}. In particular, \cite{lewis15} have found that, in the last 400 Myr, the star formation rate density of M31 has been dominated by this ring. Given that the gravitational pull is weaker at $R=10$ kpc than in the central regions of the disc, it is worth exploring a scenario in which
all the clouds are ejected, at all times, from a ring located between 10 and 12 kpc in the disc of M31. Note that this is an extreme scenario (and therefore a conservative test), as the SFR ring is likely to rather be a relatively recent and short-lived feature, as emphasised by the young ages probed by the studies above. We performed a nested sampling analysis over the same parameter space and using the same likelihood and priors adopted for the fiducial model. The results of this test are reported in the top part of Table~\ref{tab:tests}, where we can see how the values of the best-fit parameters are well within the uncertainties of what we found with our previous analysis, leading to a supernova efficiency comparable with the value found with our fiducial model. We therefore conclude that deviations from the star formation density profile adopted in this work do not affect our conclusions.\\ 
Finally, in addition to the star formation rate, a central AGN could also have an impact on the surrounding CGM. \cite{zhang19agn} found, using X-ray spectroscopy, a possible evidence in M31 for a past AGN event that happened half a million years ago. In this very short timescale, even if the clouds were ejected at $2000\ \rm{km}\ \rm{s}^{-1}$ they would reach a distance of about 1 kpc from the centre, much smaller than the distances that we are probing in this study.
We cannot exclude that previous events of activity could have also played a role. However, modelling these previous events would require numerous poorly constrained assumptions on the duty cycle, energetics and region of influence of AGN feedback for M31. This is beyond the scope of this work and is left for future studies.\\
Based on our results of Section~\ref{resultsOut} and on the above considerations, we conclude that the majority of the cool CGM of M31 is not produced by a supernova-driven galactic wind.  As mentioned in Section~\ref{resultsOut}, this in line with the findings of \citetalias{afruni20}, where we obtained a similar result for a sample of about 40 nearby star-forming galaxies selected from the COS-Halos and COS-GASS samples \citep[][]{werk12,borthakur15}. These results are also consistent with what found by \cite{fielding20smaug} in idealized hydrodynamical simulations, where feedback from the central galaxy is not able to bring cool gas in the outer CGM.\\
Galactic outflows could however have some role in the production of cold gas in the inner parts of the halo, a few kpc above the disc \citep[e.g.][]{kim18}, at distances that are not probed by the AMIGA and the COS-Halos/COS-GASS data. At these heights, the cold/hot gas interface might also lead to condensation of the coronal gas \citep[e.g.][]{marinacci10accr,armillotta16,gronke18,gronnow18,kooij21} and accretion of cold gas onto the galaxy, in a galactic fountain process \citep[][]{marasco13,fraternali17}. The study of the inner layers of the CGM is outside the scope of this work.
\subsection{Inflow}\label{Infdisc}
\subsubsection{Origin of the cool gas}
Our preferred scenario for the origin of the cool CGM of M31 is instead direct cold accretion of external gas into its halo. The results outlined in Section~\ref{resultsIn} are entirely consistent with a picture where the infalling cool clouds are coming from the cosmological accretion of gas into the halo.\\ 
We find an average cool gas mass accretion of $15.3\ M_{\odot} \rm{yr}^{-1}$. This is slightly higher than the average accretion rate expected for a halo of $M_{\rm{vir}}=2\times10^{12}\ M_{\odot}$ at $z=0$ ($\dot{M}_{\rm{cosm}}=11.5\ M_{\odot}\ \rm{yr}^{-1}$; \citealt{correa15b}). However, the prediction is only an average value and we can expect variations for a single halo, considering differences in the environment and also taking into account the uncertainties in the virial mass of M31, as we have seen in Section~\ref{M31setup}. Moreover, the cloud infall time is approximately 2.5 Gyr and the cosmological accretion of baryons 2.5 Gyr ago is expected to be higher ($\approx15\ M_{\odot}\ \rm{yr}^{-1}$) than the current value (although this would more significantly affect the internal regions of the halo). In addition to the above considerations, part of the inflowing gas might be coming from the re-accretion of material expelled in the past by M31 or by other galaxies in and around the Local Group \citep[including M31 satellites, see for example][]{hafen19}. This could also explain the slightly higher mass accretion rate compared to the estimates of \cite{correa15a,correa15b}, which are simply based on N-body simulations. For instance, using the EAGLE cosmological hydrodynamical simulations, \cite{wright21} found that the total accretion of matter at $z=0$ (including primordial and `pre-processed' gas, material that has been previously part of a galaxy) into a halo of $2\times10^{12}\ M_{\odot}$ is $\approx25\ M_{\odot}\ \rm{yr}^{-1}$. We conclude that our estimate of $f_{\rm{accr}}$ is in line with the expectations from cosmological gas accretion.\\
We estimated a metallicity of $Z\simeq0.05\,Z_{\odot}$ (see right-hand panel of Figure~\ref{fig:Nested} and Table~\ref{tab:dynesty}), which is consistent with the fact that the IGM has been enriched, throughout the evolution of the Universe, by metals expelled from galaxies \citep[e.g.][]{vandevoort12}. Even though the metallicity of the IGM at redshift zero is uncertain, recently \cite{lehner19} have found that, at $z<1$, strong Ly$\alpha$ forest absorbers (SLFs, systems with $15<\log N_{\rm{HI}}<16.2$), which are thought to be associated with the interface between the outer CGM and the Ly$\alpha$ forest/IGM, have a unimodal metallicity distribution with a median value equal to about $0.06\,Z_{\odot}$. This is very consistent with our findings.
Our result is, however, in slight disagreement with the estimate of \cite{lehner20p}, who have found a lower limit for the metallicity of the cool CGM of M31 of $0.2\ Z_{\odot}$. 
We recall that this limit was obtained combining detected OI absorption with non-detections of HI in emission \citep[see][]{howk17}. As also explained by \cite{lehner20p}, absorption and emission measurements are characterised by inherently different spatial resolutions and it is therefore possible that their estimated metallicity 
might be affected by the beam dilution effect. Moreover, this ratio is available for only 4 components in the entire AMIGA sample and \cite{lehner20p} admit that this limit is not stringent.
We therefore do not consider our result in tension with previous estimates.\\
Finding evidence of cool accreting gas around a relatively massive star forming galaxy at redshift $z=0$ is interesting and non-trivial. In the general picture, cold gas is not expected to penetrate directly into the halos of galaxies with a virial mass $\gtrsim10^{12}\ M_{\odot}$ at $z=0$ \citep[see][]{dekel06,mandelker18}. However, the actual mode of gas accretion into the halos of galaxies is still debated and a lack of cold gas predicted by simulations may also be affected by the lack of resolution \citep[see][]{hummels19,vandevoort19}. It is possible that the cold clouds, whose motion we are describing with our models, are originated from the accreting streams of intergalactic cold gas, 
fragmented by the interactions with the pre-existing hot CGM in the outer halo. These interactions might also be responsible for the deceleration of the cool clouds, possibly explaining the preference of our models for low initial radial velocities ($\varv_{\rm{r}}\sim10\ \rm{km}\ \rm{s}^{-1}$). We therefore consider the cold accretion of gas into the halo a plausible scenario to describe the majority of the cool gas observed in the halo of the Andromeda galaxy. We emphasize however that we are describing only the accretion in the outer halo of M31 and not the direct accretion of $10^4$ K gas onto the galactic disc. A more accurate discussion on the fate of this gas is presented in the following section.
{ 
\renewcommand{\arraystretch}{1.5}
 \begin{table}
\begin{center}
\begin{tabular}{*{3}{c}}
\hline  
\hline
 Model & Parameter  & Best-fit values\\
\hline 
\multirow{5}{*}{10 kpc ring}&$\log m_{\rm{cl}}/M_{\odot}$
&7.24\raisebox{.2ex}{$\substack{+0.63 \\ -0.90}$}\\
&$\log Z/Z_{\odot}$&-1.51\raisebox{.2ex}{$\substack{+0.19 \\ -0.37}$} \\
&$\log (\varv_{\rm{kick}}/\rm{km}\ \rm{s}^{-1})$&2.77\raisebox{.2ex}{$\substack{+0.22 \\ -0.07}$} \\
&$\log \eta$&1.14\raisebox{.2ex}{$\substack{+0.32 \\ -0.40}$} \\
&$\theta/\pi$&0.44\raisebox{.2ex}{$\substack{+0.05 \\ -0.05}$} \\
\hline
\multirow{6}{*}{$M_{\rm{cor}}=0.4M_{\rm{bar}}$}&$\log m_{\rm{cl}}/M_{\odot}$& 6.75\raisebox{.2ex}{$\substack{+0.38 \\ -1.09}$} \\
&$\log Z/Z_{\odot}$&-1.40\raisebox{.2ex}{$\substack{+0.40 \\ -0.18}$}\\
&$\varv_{\rm{r}}\ (100\ \rm{km}\ \rm{s}^{-1})$& 0.11\raisebox{.2ex}{$\substack{+0.24 \\ -0.09}$}\\
&$\varv_{\rm{t}}\ (100\ \rm{km}\ \rm{s}^{-1})$& 0.13\raisebox{.2ex}{$\substack{+0.35 \\ -0.37}$}\\
&$f_{\rm{accr}}$&2.01\raisebox{.2ex}{$\substack{+0.56 \\ -0.75}$}\\
\hline
\end{tabular}
\end{center}
\captionsetup{justification=centering}
\caption[]{Results of the nested sampling analysis for the two additional models presented in Section~\ref{discussion}.}\label{tab:tests}
\end{table}
}
\subsubsection{Fate of the cool gas}\label{gasfate}
If the cool CGM of M31 is indeed the result of cosmological gas accretion into the halo, there needs to be a mechanism that prevents 15 $M_{\odot}$ of cold gas from accreting every year onto the disc of M31, which has currently a star formation of less than $1\ M_{\odot}\ \rm{yr}^{-1}$ \citep[see][]{rahmani16}. One possible explanation could be the evaporation of the infalling clouds in the hot gas.\\
Since we used an analytical approach to study the dynamics of the cool CGM, our models are limited by the absence of second-order hydrodynamical effects\footnote{In addition to the first order effects of pressure confinement and drag, which are instead included.} that could destroy the cool absorbers. High-resolution simulations \cite[e.g.][]{bruggen16,armillotta17} have shown that cold clouds travelling through a hot corona are prone to destruction primarily due to Kelvin-Helmoltz instability and thermal conduction. The latter becomes particularly important for the evaporation of clouds when the temperature of the hot ambient medium is sufficiently high, as found also through analytical arguments \citep[see][]{nipoti07}. We therefore consider the evaporation a probable fate for the cool gas penetrating the hot halo of M31. In a forthcoming paper, we plan to investigate the interactions between the different phases of the CGM using high-resolution simulations. Note that even if the clouds are destroyed before reaching the internal parts of the halo, our findings would most likely remain unaffected, since the observational data are reproduced in our model mainly by clouds that are located at large distances from the centre, as can be seen in Figure~\ref{fig:CloudDist}.\\
If a substantial part of the cool gas is evaporating into the hot ambient medium, the latter is therefore continuously increasing in mass, a behaviour that our models do not take into account. The mass of the corona that we adopted in our models is about $6\times 10^{10}\ M_{\odot}$ and an addition of up to 15 $M_{\odot}$ every year would increase the coronal mass by roughly 60\% in 2.5 Gyr (the average infall time of the clouds in our model).
To investigate what the effect of this increase would be, we performed a fit using a coronal gas with the same properties outlined in Section~\ref{model}, but twice as massive. The best-fit values of the 5 free parameters, shown in the bottom part of Table~\ref{tab:tests}, are well within the range found with our fiducial model. This result implies that the increasing mass of the hot corona is not strongly affecting our findings. Note that a more massive corona would be unphysical, as it would exceed the amount of baryons expected within the halo of M31 (since $\sim60\%$ of the baryonic mass is given by the sum of the stellar and cool CGM components).\\
We conclude therefore that our results are robust and so is our favoured interpretation that the cool CGM clouds observed in the halo of M31 are part of a large scale inflow of intergalactic gas, which is feeding the halo, but not directly the star formation in the disc of M31.
\subsection{Alternative origin}\label{origins}
As mentioned in Section~\ref{resultsIn}, the number of components per line of sight observed by the AMIGA Project is slightly larger than what predicted by our models (see Figure~\ref{fig:Comparison}). While we argued that this might be due to some simplified assumptions of our models (spherical clouds in pressure equilibrium and isotropic accretion), alternative sources for the production of the cool gas might also help to explain this small discrepancy. In addition to cosmological accretion, there are, in particular, two other main formation channels for the cool circumgalactic clouds that are not considered in this work. Although exploring in detail the impact of these processes on the CGM is beyond the scope of this work, we briefly describe them in the following.\\
The first scenario is the stripping of gas from satellite galaxies, which has been extensively observed in the local Universe \citep[e.g.][]{bruns05,grcevich09,poggianti17,johnson18,putman21} and is one of the channels for the formation of cool CGM in hydrodynamical simulations \citep[for example in galaxy groups in the EAGLE simulations;][]{marasco16}. Cold gas could be stripped from satellites and subsequently infall towards the galactic disc, in addition to the cosmological accretion. Moreover, satellites can contribute to the CGM of the main galaxy by expelling gas through winds \citep[][]{hafen19,diteodoro19}.\\
The above possibility has been explored for the AMIGA data, as some dwarf galaxies surrounding M31 (the vast majority of which are dwarf spheroidals devoid of gas) are found in close proximity of most of the lines of sight \citep{lehner20p}. Even though the overall satellite velocity distribution tends not to follow the one of the CGM absorbers, part of the observed cool gas could have been stripped from a nearby satellite galaxy because of either ram pressure or tidal stripping. This gas would then follow different trajectories with respect to the satellite, possibly explaining the differences in the velocity distribution. 
Some cold gas has probably been stripped from the disc of M33, as shown by HI observations \citep[][]{braun04,lockman12,wolfe16}, even though it seems to not affect the CGM detected by AMIGA \citep[see Appendix E of][]{lehner20p}.\\
Another possible way to produce cool CGM clouds in the halo of M31 is from the condensation of the hot coronal gas due to thermal instabilities. These would originate clumps of cold gas that will precipitate towards the centre, in addition to the cosmological accretion. The cooling time of the hot CGM is too long for spontaneous radiative cooling to happen, especially in the external parts due to the very low densities, therefore some strong perturbation (such as turbulence, see for example \citealt{voit18}) is needed to create dense pockets of gas with short cooling times, which could then perhaps trigger a thermal instability. These perturbations could also be caused by external processes like the cosmological accretion of IGM and/or satellite galaxies \citep[][]{nelson20,esmerian21}.
However, whether or not these instabilities can develop at all is still under debate \citep[see][]{binney09,mccourt12,sharma12,nipoti13,nipoti14,sormani19}.
\section{Summary and conclusions}\label{conclusions}
With this study, we have investigated the properties of the cool CGM residing in the halo of the Andromeda galaxy (M31), using semi-analytic parametric models, where we describe the circumgalactic gas as a flow of cool clouds embedded in a hot ambient medium (corona). Our models take into account the effect of gravity and of the hydrodynamical interactions (drag and pressure equilibrium) of the clouds with the corona, which also has an angular momentum motivated by cosmological prescriptions. We have compared our model predictions with the observational data of the AMIGA project \citep[][]{lehner20p}. Our goal was to reproduce the kinematics, the number of components and the total column densities provided by these data, in order to infer the properties of the cool CGM of M31. We investigated two different scenarios for the dynamics and origin of the cool gas: in the first scenario the clouds are part of a supernova-driven galactic wind, while in the second they are part of the cosmological accretion of gas into the galactic halo. We compared our predictions with the observations through a Bayesian analysis, which allowed us to find the best model that can successfully reproduce the data.\\
The results of our analysis are the following:
\begin{enumerate}
    \item the cool circumgalactic gas observed in the halo of M31 is not originated by supernova driven galactic winds. In order to reproduce the observations these outflows would require unphysical supernova efficiencies ($\approx700\%$) in transferring kinetic energy to the wind. Moreover, we predict that, to reproduce the AMIGA data, the outflowing gas would need to have low metallicities, in contrast
    with the expectation for material ejected from the central galaxy. We therefore discard this scenario as a viable way to describe the observations;
    \item the covering factors, kinematics and column densities of the absorbers in the AMIGA sample are well reproduced by our inflow models. In particular, in these models the gas clouds have a mass of about $5\times10^6\ M_{\odot}$ and start their infall from the virial radius with low radial and tangential velocities. They then fall towards the centre with a spiraling motion, while they exchange angular momentum with a surrounding hot halo.
    We find a cool gas accretion rate at the virial radius of $\approx15\ M_{\odot}\ \rm{yr}^{-1}$, consistent with the accretion predicted by cosmological models;
    \item to reproduce the observations, the accreted material needs to have metallicities of about $0.05\,Z_{\odot}$, consistent with the low metallicity expected for gas inflowing from the intergalactic medium.
\end{enumerate}
Given our findings and conclusions, we favour a self-consistent scenario where the cool medium observed in the halo of M31 is produced by the accretion of low-metallicity intergalactic gas. Our study in fact shows how the vast majority of the observed cool gas can be described by the accretion of IGM, without the requirement of strong feedback from the central disc. Finally, we also find that most of the observed clouds, even at small projected distances from the disc, are intrinsically located in the outer parts of the halo, at large galactocentric distances. We argue that, given the second-order hydrodynamical effects that are not accounted for in this analytical analysis, these CGM clouds will likely evaporate in the hot ambient medium and therefore will not reach the disc and directly feed the star formation of M31.
\section*{Acknowledgements}
We have made extensive use of the python packages NumPy \citep{numpy} and Matplotlib \citep[][]{matplotlib}, which we would like to acknowledge. We thank the anonymous referee for a constructive report and insightful comments that helped improving the quality of this work. AA would also like to thank Pavel Mancera Pi{\~n}a for useful and helpful discussions.
GP acknowledges support from the Netherlands Research School for Astronomy (NOVA).

\vspace{-0.3cm}

\section*{Data Availability}
The data underlying this article are available in the article and in its online supplementary material.

\vspace{-0.3cm}




\bibliographystyle{mnras}
\bibliography{biblio} 



\vspace{-0.4cm}

\appendix

\section{Properties of the hot gas model}\label{hotgas}
In this Appendix we describe the properties of the hot CGM presented in Section~\ref{hotgasMod}.
We utilised the python package COROPY \citep{sormani18}. This package allows to create baroclinic models\footnote{In these models, the pressure and density are not stratified along the same surfaces and the rotational velocity is not stratified in cylinders.} of the hot rotating CGM, assuming that the pressure is stratified in ellipsoidal surfaces. The properties of such models are uniquely defined by the choice of the pressure profile along the axis of symmetry ($R=0$) of the galaxy $P_{\rm{axis}}(z)$ and by the axis ratio of the isobaric ellipsoidal surfaces $b/a=q(a)$ as a function of the ellipse semi-major axis. In this work, $P_{\rm{axis}}(z)$ is obtained solving the equation of hydrostatic equilibrium along the $R=0$ axis and assuming a polytropic distribution along this axis: $P_{\rm{axis}}(z)=[\rho_{\rm{axis}}(z)]^{\gamma}$, resulting in
\begin{ceqn}
\begin{equation}\label{eq:pressure}
P_{\rm{axis}}(z)= P_0 \left[ 1 + \frac{\gamma-1}{\gamma}\frac{\mu m_{\rm{p}}}{kT_0}\left(\Phi_{\rm{axis}}(z) - \Phi_{\rm{axis},0} \right)\right]^{\gamma/(\gamma-1)},
\end{equation}
\end{ceqn}
where $m_p$ and $k$ are respectively the proton mass and the Boltzmann constant,
$\gamma$ is the politropic index, while $T_0$ and $P_0$ are temperature and pressure at some reference height $z_0$. Here we adopt $\gamma=1.2$ and $T_0=1.3\times 10^6$\ K at $z_0=100$ kpc, with $P_0=P_{\rm{axis}}(z_0)=n_0kT_0$, with the value of $n_0$ defined below. Finally, $\Phi_{\rm{axis}}(z)$ is the potential at the axis of symmetry, with $\Phi_{\rm{axis,0}}=\Phi_{\rm{axis}}(z_0)$. Only in this part of the model, we simplified the potential to just the NFW component. This is because, using the full potential, the assumption of \cite{sormani18} that the pressure is stratified on ellipses would break the reality condition for the rotation velocity \citep[e.g.][]{barnabe06}. However, this effect is limited to only a few kpc above the disc plane and is unimportant for this work.
For the axis ratio we use a hyperbolic function:
\begin{ceqn}
\begin{equation}\label{eq:q}
q(r)= \frac{q_{\rm{min}}r_{\rm{min}}+q_{\rm{max}}r}{r_{\rm{min}}+r}\ ,
\end{equation}
\end{ceqn}
with $r_{\rm{min}}=5$ kpc, $q_{\rm{min}}=0.6$ and $q_{\rm{max}}=0.9999$.
   \begin{figure}
   \includegraphics[clip, trim={0.5cm 0.1cm 1.5cm 1cm}, width=\linewidth]{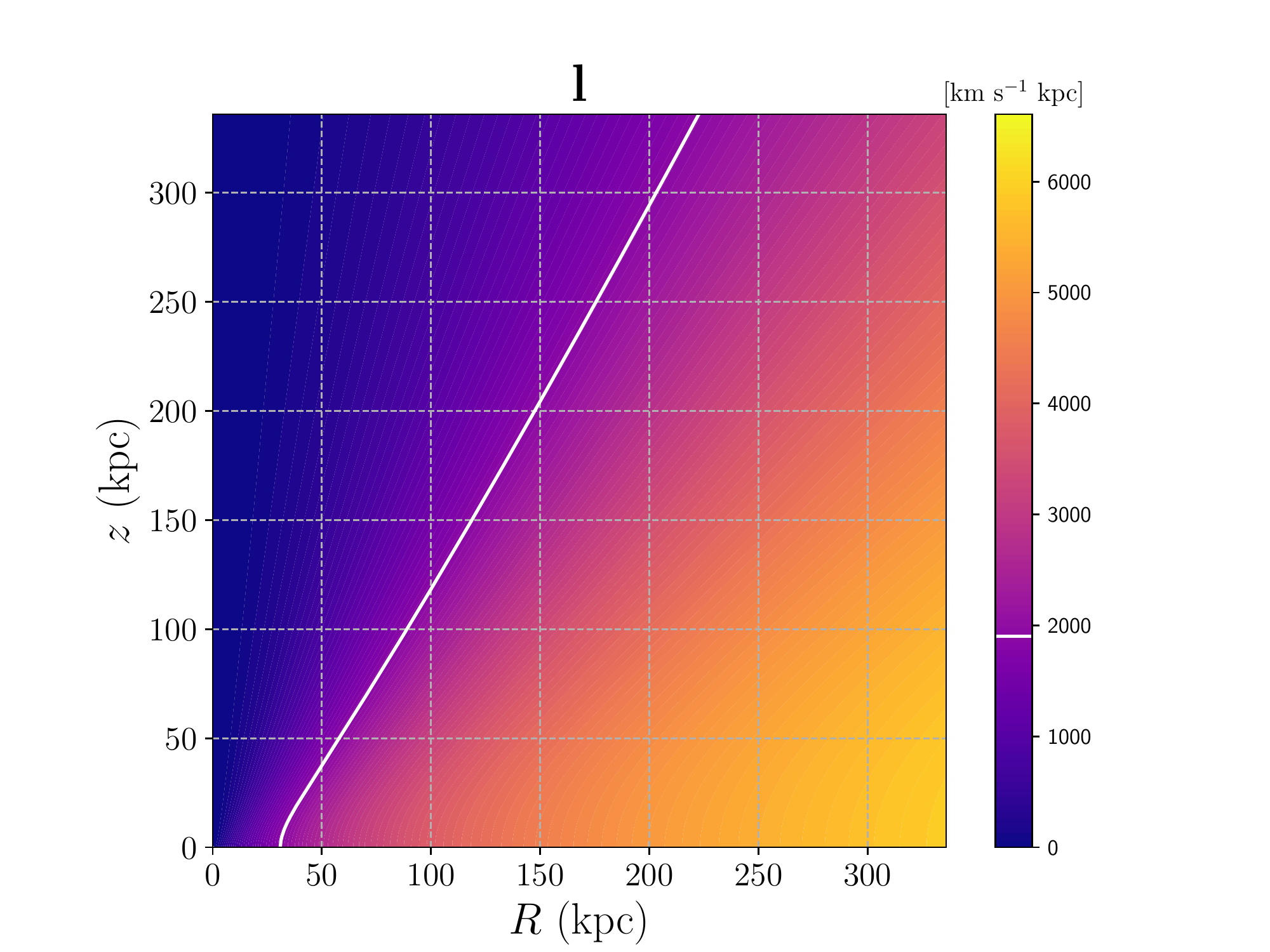}
   \caption{Specific angular momentum in our model of the hot CGM of M31 (see also Section~\ref{hotgasMod}), as a function of the cylindrical radius $R$ and height $z$. The white contour indicates as a reference the angular momentum of the stellar component of M31.}
              \label{fig:angmomHot}%
    \end{figure}\\
All the parameters in equations~\eqref{eq:pressure} and \eqref{eq:q} ($\gamma, T_0, r_{\rm{min}}, q_{\rm{min}}, q_{\rm{max}}$) have been chosen (by trial and error) in order to have a realistic model for the corona of M31, in particular to reproduce an angular momentum in agreement with the theoretical expectations (for more details see Section~\ref{hotgasMod}).
Once all the other parameters are fixed, the normalization factor $n_0$ is chosen in order to have a total mass in the hot CGM phase equal to 20\% of the total baryonic mass expected within the halo, inferred by multiplying the virial mass by the cosmological baryon fraction 0.158 \citep{planck20}. This corresponds to a mass of the corona of about $6\times 10^{10}\ M_{\odot}$.\\
With the choice of parameters reported above, we find an average value of the specific angular momentum in our corona model equal to $3207\  \rm{km}\ \rm{s}^{-1}\ \rm{kpc}$, which is 20$\%$ higher than the DM estimate (see Section~\ref{hotgasMod}). This difference could be attributed to the loss of low angular momentum material ejected by feedback in the past, consistent with the fact that the corona adopted in this work accounts for only 20$\%$ of the baryons expected within the dark matter halo \citep[e.g.][]{pezzulli17}. The map of the specific angular momentum given by our final model is shown in Figure~\ref{fig:angmomHot}, with the white contour showing $l=1900\ \rm{km}\ \rm{s}^{-1}\ \rm{kpc}$, giving an indication of where the angular momentum of the corona equals that of the disc (the lowest portion of this line could be slightly shifted inwards if including the disc contribution to the gravitational potential). Note how the specific angular momentum increases with the cylindrical radius $R$, as expected for a realistic model where the highest angular momentum material was accreted at later epochs and deposited in the outer regions.

\bsp	
\label{lastpage}
\end{document}